\documentclass[aps,prd,showpacs,twocolumn,superscriptaddress,nofootinbib,preprintnumbers]{revtex4-1} 

\makeatletter
\def\p@subsection{}
\makeatother

\usepackage{graphicx,amsmath,amssymb,lipsum,bm}
\usepackage[usenames,dvipsnames]{xcolor}
\definecolor{xlinkcolor}{rgb}{0.7752941176470588, 0.22078431372549023, 0.2262745098039215}

\usepackage[colorlinks=true,citecolor=xlinkcolor,linkcolor=xlinkcolor,urlcolor=xlinkcolor, backref=false,pdfborder={0 0 0}]{hyperref}

\usepackage[normalem]{ulem}
\usepackage[utf8]{inputenc} 
\usepackage{mathtools}
\usepackage{aas_macros}
\usepackage{float}

\usepackage{multirow}
\usepackage{ulem}
\usepackage{diagbox}

\usepackage[dvipsnames]{xcolor}
\usepackage{mathrsfs}

\newcommand{\be}{\begin{equation}}
\newcommand{\ee}{\end{equation}}
\newcommand{\beqa}{\begin{eqnarray}}
\newcommand{\eeqa}{\end{eqnarray}}

\newcommand{\bseq}{\begin{subequations}}
\newcommand{\eseq}{\end{subequations}}

\renewcommand{\ln}{\mathop{\rm ln}\nolimits}

\newcommand{\Mpc}{\,{\rm Mpc}}

\newcommand{\Trh}{T_{\rm rh}}
\newcommand{\fend}{f_{\rm end}}
\newcommand{\frh}{f_{\rm rh}}
\newcommand{\fpta}{f_{\rm PTA}}
\newcommand{\flvk}{f_{\rm LVK}}
\newcommand{\Hz}{{\rm Hz}}
\newcommand{\eV}{{\rm eV}}
\newcommand{\GeV}{{\rm GeV}}

\def\gsim{\raise0.3ex\hbox{$\;>$\kern-0.75em\raise-1.1ex\hbox{$\sim\;$}}}
\def\lsim{\raise0.3ex\hbox{$\;<$\kern-0.75em\raise-1.1ex\hbox{$\sim\;$}}}

\def\beqn#1{\begin{equation}\label{#1}}
\def\eeqn{\end{equation}}

\def\beqa#1{\begin{eqnarray}\label{#1}}
\def\eeqa{\end{eqnarray}}

\newcommand{\delf}[1]{\textcolor{cyan}{#1}}

\begin{document}

\title{Inflationary interpretation of the gravitational-wave signal \\
in the European Pulsar Timing Array DR2 with constraints}

\author{Philippe Turgeon}
\email{pxt561@student.bham.ac.uk}
\affiliation{School of Physics and Astronomy,\\
University of Birmingham, Edgbaston, Birmingham B15 2TT, United Kingdom}

\author{Chiara Caprini}
\email{chiara.caprini@unige.ch}
\affiliation{Theoretical Physics Department, CERN, CH-1211 Gen\`eve, Switzerland}
\affiliation{D\'epartement de Physique Th\'eorique and Center for Astroparticle Physics,\\
Universit\'e de Gen\`eve, 24 quai Ernest  Ansermet, 1211 Gen\`eve 4, Switzerland}

\author{Anton~Chudaykin}
\email{anton.chudaykin@unige.ch}
\affiliation{D\'epartement de Physique Th\'eorique and Center for Astroparticle Physics,\\
Universit\'e de Gen\`eve, 24 quai Ernest  Ansermet, 1211 Gen\`eve 4, Switzerland}

\author{Martin Kunz}
\email{martin.kunz@unige.ch}
\affiliation{D\'epartement de Physique Th\'eorique and Center for Astroparticle Physics,\\
Universit\'e de Gen\`eve, 24 quai Ernest  Ansermet, 1211 Gen\`eve 4, Switzerland}

\author{Delphine Perrodin}
\email{delphine.perrodin@inaf.it}
\affiliation{INAF - Osservatorio Astronomico di Cagliari, via della Scienza 5, 09047 Selargius (CA), Italy}

\author{I.~Cognard}
\email{ismael.cognard@cnrs-orleans.fr}
\affiliation{LPC2E, OSUC, Univ Orleans, CNRS, CNES, Observatoire de Paris, F-45071 Orleans, France}
\affiliation{ORN, Observatoire de Paris, Universit\'e PSL, Univ Orleans, CNRS, 18330 Nan\c{c}ay, France}

\author{L.~Guillemot}
\email{lucas.guillemot@cnrs-orleans.fr}
\affiliation{LPC2E, OSUC, Univ Orleans, CNRS, CNES, Observatoire de Paris, F-45071 Orleans, France}
\affiliation{ORN, Observatoire de Paris, Universit\'e PSL, Univ Orleans, CNRS, 18330 Nan\c{c}ay, France}

\author{G.~Theureau}
\email{gilles.theureau@cnrs-orleans.fr}
\affiliation{LPC2E, OSUC, Univ Orleans, CNRS, CNES, Observatoire de Paris, F-45071 Orleans, France}
\affiliation{ORN, Observatoire de Paris, Universit\'e PSL, Univ Orleans, CNRS, 18330 Nan\c{c}ay, France}

\begin{abstract}
The second data release of the European Pulsar Timing Array (EPTA) collaboration provides evidence for the presence of a gravitational-wave (GW) background. In this work, we explore a potential cosmological interpretation of this signal in terms of inflationary scenarios.
We parametrize the tensor power spectrum in terms of the tensor-to-scalar ratio $r$, the tensor spectral index $n_t$, the reheating temperature $T_{\rm rh}$ and the cut-off frequency $\fend$. 
We incorporate all relevant observational constraints, including those from Cosmic Microwave Background, Big Bang Nucleosynthesis and LIGO–Virgo–KAGRA observations.
We demonstrate that imposing these constraints consistently reduces the region of parameter space that provides a viable interpretation of the EPTA signal, to $-11.66 \lesssim \log_{10}r \lesssim -1.45$, $1.32 \lesssim n_t \lesssim  2.47$, $1.78  \,{\rm MeV} \lesssim \Trh \lesssim 28.2 \,{\rm GeV}$ and $75.86\,{\rm n\Hz} \lesssim \fend \lesssim 14.45\, {\rm \Hz}$ at the $95\%$ confidence level.
This favours the scenario in which the GW spectrum in the EPTA frequency band originates from tensor modes that re-entered the Hubble radius during the radiation-dominated era, allowing for a higher $r$ and a flatter spectrum. 
However, $\Trh$ must take very low values, which are challenging to explain theoretically.

\end{abstract}
\maketitle
\section{Introduction}\label{sec:intro}

In 1978-1979, Sazhin \cite{sazhin_opportunities_1978} and Dettweiler \cite{detweiler_pulsar_1979} proposed using residuals \delf{}{from} the Times-Of-Arrival (TOAs) of pulsar signals to detect a stochastic Gravitational Wave Background (GWB). 
Foster and Backer \cite{1990ApJ...361..300F} were the first to propose monitoring highly stable millisecond pulsars in the search for a GWB, an approach now known as Pulsar Timing Arrays (PTAs).
By jointly modeling GW signals and pulsar noise, this technique led to the detection of a common red signal (CRS) across multiple pulsars, which, if interpreted as a GWB, would be compatible with astrophysical sources.

In fact, after many years of progressively tightening upper limits on the GWB amplitude, the North American Nanohertz Observatory for Gravitational Waves (NANOGrav)~\cite{McLaughlin:2013ira} reported in 2020 the detection of a long-term, low-frequency stochastic signal with common spectral properties across the pulsar array ~\cite{NANOGrav:2020bcs}. This finding was subsequently confirmed by the European Pulsar Timing Array (EPTA)~\cite{EPTA:2016ndq} in~\cite{EPTA:2021crs} and by the Parkes Pulsar Timing Array (PPTA)~\cite{Manchester:2012za} in~\cite{Goncharov:2021oub}. The International Pulsar Timing Array (IPTA) consortium, which combines data from several PTAs, also confirmed this observation~\cite{Antoniadis:2022pcn}. Despite the detection of a common red signal to a high significance, the latter may originate from intrinsic pulsar processes or from a common systematic noise (e.g. clock errors). Definitive evidence for a GW origin of the observed signal can be established through the measurement of inter-pulsar correlations following the Hellings-Downs (HD) angular pattern~\cite{Hellings:1983fr}. 
The most recent results from NANOGrav and EPTA provide compelling evidence for HD correlations at the $3\text-4\sigma$ level~\cite{NANOGrav:2023gor,EPTA:2023fyk}. 
The observed HD-correlated signal is consistent with the one expected from a cosmic population of supermassive black hole binaries (SMBHBs), {but the observed best fit spectrum is significantly flatter}~\cite{EPTA:2023xxk}. 

Although SMBHBs are the most widely considered explanation for the observed GWB \cite{NANOGrav:2023gor,EPTA:2023fyk}, more exotic explanations are possible \cite{NANOGrav:2023hvm,EPTA:2023fyk,Figueroa:2023zhu}, including cosmic inflation~(see e.g.~\cite{Guzzetti:2016mkm,Kuroyanagi:2020sfw,Vagnozzi:2023lwo}), scalar-induced GWs~(see e.g.~\cite{Domenech:2021ztg,Yuan:2021qgz,Balaji:2023ehk,Franciolini:2023pbf}, first-order phase transitions~(see e.g.~\cite{Caprini:2019egz,Hindmarsh:2020hop,RoperPol:2022iel}), cosmic strings and domain walls~(see e.g.~\cite{Vilenkin:1984ib,Hindmarsh:1994re,Saikawa:2017hiv,EuropeanPulsarTimingArray:2023lqe}). 
Cosmological GWBs generated in the early Universe are stochastic in nature and can therefore reproduce the angular spatial correlations described by the HD curve~(for demonstrations on how GW signals of stochastic nature lead to the HD correlation, see e.g.~\cite{Cornish:2013aba,Maggiore:2007ulw}). 
In this work, we explore a cosmological interpretation of the GWB signal as arising from inflation. 

In the standard scenario of inflation, tensor quantum vacuum fluctuations of the metric are amplified by the accelerated expansion. 
The tensor modes subsequently re-enter the Hubble radius, leading to an inflationary GWB (IGWB). 
This occurs during the full cosmological timeline, including reheating, radiation, and matter domination.
The detection of an IGWB would provide valuable insight into the inflationary model and the subsequent reheating phase.

The impact of IGWs on the Cosmic Microwave Background (CMB) is commonly quantified by the tensor-to-scalar ratio, $r$, and the tensor spectral index $n_t$ of the tensor perturbations. In the simplest inflationary scenarios -- single-field slow-roll inflation -- these two parameters are linked via the consistency relation, $n_t=-r/8$ \cite{Planck:2013jfk}. Given that $r$ is tightly constrained from above by current CMB observations, $r<0.034$ at $95\%$ CL~\cite{BICEP:2021xfz,Balkenhol:2025wms},
this implies $n_t\approx0$, causing the spectral energy density of this signal to be out of reach in the frequency band of current and planned GW detectors, including PTAs. However, this relation only applies to minimal single-field slow-roll inflation and can be avoided in extended inflationary scenarios. In particular, a blue-tilted spectrum can be produced in several non-minimal inflationary models~ (see e.g.~\cite{Anber:2012du,Cook:2011hg,Namba:2015gja,Dimastrogiovanni:2016fuu,Caldwell:2017chz,Piao:2004tq,Kobayashi:2010cm,Endlich:2012pz,Fujita:2018ehq}).

In this work, we adopt a model-independent approach and parametrize the primordial tensor power spectrum in terms of the parameters $r$ and $n_t$ without imposing the slow-roll consistency relation between them, and instead vary both parameters independently over broad prior ranges. 
In addition, we sample the reheating temperature $T_{\rm rh}$, which marks the end of the reheating period, and the inflationary cutoff of the IGW spectrum $\fend$, which marks the end of inflation.
This model-independent approach, already adopted by the NANOGrav~\cite{NANOGrav:2023hvm} and EPTA~\cite{EPTA:2023xxk} collaborations, allows us to derive model-independent insight on the features that inflationary scenarios should possess in order to explain PTA measurements. 

The present work extends the NANOGrav and EPTA analyses in two main directions.
First, we implement the observational constraints on the amplitude of the GWB from Big Bang Nucleosynthesis (BBN), CMB and LIGO–Virgo–KAGRA (LVK) directly within the Markov Chain Monte Carlo (MCMC) sampling, whereas earlier studies either did not incorporate these bounds~\cite{EPTA:2023xxk}, or incorporated them {\it a posteriori}, after processing the chains~\cite{NANOGrav:2023hvm}. 
This approach allows us to quantify the impact of these observational constraints directly on the posterior distributions by consistently propagating their effects into the constraints on cosmological parameters: as we will see, this leads to somewhat different conclusions with respect to those drawn in \cite{NANOGrav:2023hvm,EPTA:2023xxk}.

Second, we adopt a more flexible theoretical framework compared to previous analyses~\cite{NANOGrav:2023hvm,EPTA:2023xxk}. Our model incorporates an additional free parameter, the end-point of the IGW spectrum, $\fend$, which marks the end of inflation and the onset of reheating. Allowing $\fend$ to vary freely accounts for a prolonged reheating phase~(see e.g.~\cite{Allahverdi:2010xz,Kuroyanagi:2014qza,Kuroyanagi:2010mm}).
In contrast, earlier studies within the PTA Collaborations employed simpler models: they either did not model the end of inflation and reheating~\cite{EPTA:2023xxk}, or set the endpoint of the IGW spectrum to a constant value corresponding to the maximum amount of e-folds of reheating, for a given $T_{\text{rh}}$, which avoids the violation of LVK + BBN bounds~\cite{NANOGrav:2023hvm}. In our analysis, we demonstrate that sampling over $\fend$ in combination with observational bounds provides constraints on this parameter.

In this work, we use the second EPTA data release DR2new, consisting of 10.3 years of timing residual observations of 25 millisecond pulsars with the five largest radio telescopes in Europe: the Effelsberg telescope (Germany), the Westerbork Synthesis Radio Telescope (the Netherlands), the Lovell telescope at Jodrell Bank Observatory (UK), the Nan{\c c}ay Radio Telescope (France) and the Sardinia Radio Telescope (Italy), as well as the same telescopes being used simultaneously as the Large European Array for Pulsars (LEAP). Assuming an inflationary origin of the PTA signal, we derive constraints on the aforementioned parameters from the EPTA data alone and in combination with relevant physical 
constraints, including CMB, BBN and LVK observations. For ease of calculations, we use the Common Uncorrelated Red Noise (CURN) framework for our analysis; we assume that, considering the large uncertainties on the parameters of the observed signal, the cosmological GWB is well described by the CURN signal without necessarily requiring HD correlations in the calculations.
We find that our 
phenomenological IGWB model provides a viable interpretation of the EPTA signal, and we identify a region of parameter space consistent with current observations. 

The remainder of this paper is structured as follows.
In Sec.~\ref{sec:model} we present our theoretical model.
In Sec.~\ref{sec:analys}, we describe the datasets employed in this work, including EPTA data, external physical constraints from the CMB, BBN, LVK observations, as well as our analysis pipeline. 
Our main cosmological results are presented in Sec.~\ref{sec:res}.
In Sec.~\ref{sec:comparison} we compare our findings with those of EPTA~\cite{EPTA:2023xxk} and NANOGrav~\cite{NANOGrav:2023hvm}. 
We conclude in Sec.~\ref{sec:conc}.
In Appendix~\ref{sec:bayes}, we perform the analysis when sampling over Solar System Ephemeris (SSE) parameters using the \texttt{BAYESEPHEM} model \cite{NANOGrav:2020tig}, as opposed to the SSE model DE440 \cite{Park_2021} used throughout this analysis. 

\section{GWB model}\label{sec:model}

We consider a four-parameter framework characterized by the tensor-to-scalar ratio $r$, the tensor spectral index $n_t$, the reheating temperature $\Trh$ and the high-frequency cutoff of the IGW spectrum, $\fend$. 
In our analysis, we assume a constant $n_t$ over the entire frequency range, from CMB scales to those corresponding to PTAs and ground-based GW detectors. 
Following \cite{Kuroyanagi:2014nba, Kuroyanagi:2020sfw} (see also \cite{NANOGrav:2023gor,Caprini:2018mtu}), the IGW spectrum is modeled as\footnote{Note that we average the Bessel function $\lim_{k\eta_0\to \infty} j_1^2(k\eta_0)= 1/(2k\eta_0)^2$ for modes well inside the horizon today.}
\begin{eqnarray}\label{eq:OmGW}
    \lefteqn{\Omega_{\text{GW}}(f) =} \nonumber \\ & \frac{3}{32\pi^2}\left(\frac{f}{H_0}\right)^2\left(\frac{1}{f\eta_0}\right)^4rA_{\text{s}}\left(\frac{f}{f_{\text{CMB}}}\right)^{n_t}\mathcal{T}_*^2(f)\,,
\end{eqnarray}
where $f=k/(2\pi a_0)$ denotes the GW frequency today with $k$ comoving momentum, $H_0$ is the Hubble factor today, $\eta_0 \simeq 2/[H_0\left(\sqrt{\Omega_\text{m} + \Omega_\text{r}} + \sqrt{\Omega_{\text{r}}}\right)]$ is the current conformal time with $\Omega_\text{r}$ and $\Omega_\text{m}$ the present-day radiation and matter energy density fraction, respectively, and $A_\text{s} = 2.1 \times 10^{-9}$ is the amplitude of the scalar perturbations defined at the CMB pivot scale $f_{\rm CMB} = 0.05 \Mpc^{-1}/(2\pi) \simeq 7.73 \times 10^{-17}\,\Hz$ \cite{Planck:2018vyg}.
The transfer function $\mathcal{T}_*^2(f)$ encodes the standard cosmological evolution of GWs after horizon re-entry and is also derived in \cite{Kuroyanagi:2014nba,Kuroyanagi:2020sfw}:
\begin{multline}\label{eq:tran}
    \mathcal{T}_*^2(f,T_{\text{rh}},\fend) = \Omega_{\text{m}}^2\mathcal{T}_1^2(f)\mathcal{T}^2_2(f,T_{\text{rh}}) \\
    \times\left(\frac{g_{*,s}^0}{g_{*,s}(f)}\right)^{4/3}\left(\frac{g_*(f)}{g_*^0}\right)\Theta(f - \fend)
\end{multline}
\begin{equation}\label{eq:T1}
  \mathcal{T}^2_1(f) = 1 + 1.57\left(\frac{f}{f_\text{eq}}\right) + 3.42\left(\frac{f}{f_{\text{eq}}} \right)^2
 \end{equation}
\begin{multline}\label{eq:T2}
    \mathcal{T}^2_2(f,T_{\text{rh}})= \\ \left(1 - 0.22\left(\frac{f}{\frh(\Trh)} \right)^{1.5} + 
    0.65\left(\frac{f}{\frh(\Trh)} \right)^2\right)^{-1}
\end{multline}
Here $\mathcal{T}_1(f)$ and $\mathcal{T}_2(f,T_{\text{rh}})$ are fitting functions, encoding the changes in the IGWB spectral shape due to the stage at which the modes re-enter the Hubble radius. In particular, the function $\mathcal{T}_1(f)$ connects the GW spectrum of modes entering the Hubble radius before and after matter-radiation equality, and has first been obtained in~\cite{Turner:1993vb}. 
It depends on $f_{\text{eq}} = H_0\Omega_{\rm m} /(\pi\sqrt{2\,\Omega_{\rm r}})\sim
2.1 \times 10^{-17}\,\Hz$, the frequency of modes re-entering the Hubble radius at matter–radiation equality~\cite{Caprini:2018mtu}. 

The function $\mathcal{T}_2(f,T_{\text{rh}})$ describes the modifications of the spectral shape induced by reheating~\cite{Kuroyanagi:2020sfw}. 
The effects of reheating were modeled in~\cite{Kuroyanagi:2008ye,Kuroyanagi:2020sfw} by numerically solving the equations describing the perturbative decay of the Inflaton, leading to a matter-dominated era, while accounting for several relevant effects (e.g., changes of the effective number of relativistic degrees of freedom and the anisotropic stress of free-streaming neutrinos)~\cite{Kuroyanagi:2008ye}. 
The computation assumes a quadratic potential for the scalar field, which is in general appropriate close to the minimum of the potential.\footnote{Note, however, that the shape of the generated IGWB spectrum during reheating depends on the choice of the Inflaton potential,
see~\cite{Kuroyanagi:2008ye} for details.}
$\mathcal{T}_2(f,T_{\text{rh}})$ depends on $\frh$, the frequency of modes entering the Hubble radius at the end of reheating: 
\begin{eqnarray}\label{eq:frh}
\frh(\Trh) &=&\frac{1}{2\pi}\frac{a_{\rm rh}H_{\rm rh}}{a_0}\\ 
&\simeq& 7.8 \times10^{-9}\,\Hz\,
\bigg(\frac{g_{*,s}^{0}}{g_{*,s}^{\rm rh}}\bigg)^{1/3}\sqrt{g_{*}^{\rm rh}}
\left(\frac{\Trh}{1\,\GeV}\right).\nonumber
\end{eqnarray} 
The effective number of relativistic degrees of freedom contributing to the radiation energy density and radiation entropy are $g_{*}$ and $g_{*,s}$, respectively. Note that $g_{*,s}=g_*$ before neutrino decoupling.
The superscript ``0'' denotes their values at the present time, where $g_*^0=g_{*,s}^0=2$.
The values of the relativistic degrees of freedom depend on the temperature of the hot plasma at the time when the mode with frequency $f$ re-enters the Hubble radius. We adopt the temperature dependence of $g_{*}(T)$ and $g_{*,s}(T)$ from~\cite{Laine:2015kra}, and then convert the temperature to frequency using the relation at the time of horizon reentry $f=k_*/(2\pi a_0)= a(T)H(T)/(2\pi)$. 
To avoid clutter, we refer to these functions simply as $g_{*}(f)$ and $g_{*,s}(f)$.

The additional factor $\Omega_{\rm m}^2$ in Eq.~\eqref{eq:tran} accounts for the present-day matter abundance in the $\Lambda$CDM model, which suppresses the present GW energy density~\cite{Turner:1993vb,Caprini:2018mtu}.
Furthermore, the Heaviside function $\Theta(f - \fend)$ imposes a sharp cut-off on modes that are smaller than the Hubble scale at the end of inflation. 
The GW modes that are sub-Hubble  at the end of inflation, $f > \fend$, are not amplified and therefore contribute negligibly to the IGWB power spectrum.
$\fend$ depends on the Hubble parameter at the end of inflation $H_{\rm end}$ \cite{NANOGrav:2023gor}, 
\begin{eqnarray}\label{eq:fend_formula}
    \fend &=&\frac{1}{2\pi}\frac{a_{\rm end}}{a_{\rm rh}}\frac{a_{\rm rh}}{a_0}H_{\rm end}\\
    &=&\frac{1}{2\pi} \bigg(\frac{g_{*,s}^{0}}{g_{*,s}^{\rm rh}}\bigg)^{1/3} \left(\frac{\pi^2}{90}g_{*}^{\rm rh}\right)^{1/3}\left(\frac{ T_{\rm rh} H_{\rm end}}{M_{\rm Pl}^2}\right)^{1/3}T_0\,,\nonumber
\end{eqnarray}
where $M_{\rm Pl}$ is the reduced Planck mass. 
An upper bound on the scale of inflation $H_{\rm inf}$ can be inferred from the measurement of the scalar spectrum amplitude $A_\text{s}=2.1\times 10^{-9}$ combined with the upper bound on $r$ from CMB B-mode polarization, see the first line in Eqs.~\eqref{eq:cmb}:
\begin{equation}
\label{eq:Hinf}
    H_{\rm inf} \lesssim \frac{\pi}{\sqrt{2}} M_{\rm Pl} \sqrt{r A_\text{s}}\simeq 6.8\times 10^{13}\,{\rm GeV}\,.
\end{equation}
Since $H_{\rm end}\leq H_{\rm inf}$, from Eq.~\eqref{eq:Hinf} we can infer an upper bound on $\fend$ from the scenario of instantaneous reheating, which corresponds to maximal $T_{\rm rh}$:
\begin{eqnarray}\label{eq:fend_value}
        \fend &\leq & \frac{1}{2\sqrt{\pi}} \bigg(\frac{g_{*,s}^{0}}{g_{*,s}^{\rm rh}}\bigg)^{1/3} \left(\frac{g_{*}^{\rm rh}}{90}\right)^{1/4}\sqrt{\frac{ H_{\rm end}}{M_{\rm Pl}}}\,T_0\\
        &\lesssim & 1.5\times 10^8 \,{\rm Hz}\,,\nonumber
\end{eqnarray}
where in the second line we have indicatively substituted the canonical value $g_{*,s}^{\rm rh}=g_{*}^{\rm rh}=106.75$, i.e.~the Standard Model value at the EWPT. 
Therefore, in the minimal scenario with slow roll inflation and instantaneous reheating, $\fend$ is much higher than the PTA frequency band.

The general behaviour of the full transfer function is the following:
for modes re-entering the Hubble radius during the matter-dominated and reheating epochs, the transfer function is frequency-independent, $\mathcal{T}_*^2 \propto f^0$, while for modes re-entering the Hubble during the radiation-dominated era, it scales as $\mathcal{T}_*^2 \propto f^2$. 
The product of the fitting functions $\mathcal{T}_1$ and $\mathcal{T}_2$ provides a smooth and continuous evolution throughout the  cosmological history, while also accounting for the impact of reheating on the IGWB spectrum~\cite{Kuroyanagi:2020sfw}. It interpolates the analytical solutions that can be derived for modes re-entering the horizon deep in the matter- or radiation-dominated stages of the evolution of the Universe~\cite{Caprini:2018mtu}. 
The model of $\Omega_{\text{GW}}(f)$ is then converted in terms of the spectral density $S(f)$ through the relation
\begin{equation}
    S(f) = \frac{\Omega_{\text{GW}}(f)H_0^2}{8\pi^4f^5}\,,
    \label{eq:S}
\end{equation}
used by EPTA \cite{EPTA:2023xxk} to describe the common red noise spectrum.

\section{Data and analysis}\label{sec:analys}

\subsection{EPTA data}\label{sec:analys1}

We use the EPTA 25-pulsar dataset, \texttt{DR2new}, collected over 10.3 years of observations~\cite{EPTA:2023fyk}. These measurements were obtained with new-generation backends that provide substantially wider bandwidths and improved sensitivity, yielding robust evidence for a GWB signal with Hellings–Downs correlations ($\gtrsim 3\sigma$ significance). We do not use the full DR2 dataset because evidence of HD quadrupolar correlation of the common process is weaker in that dataset ($<2\sigma$ significance), potentially due to the lower quality of early data collected with narrowband backends; see~\cite{EPTA:2023fyk} for a detailed discussion. 

The EPTA measurements in Fourier space are discretised in the first nine frequency bins, $f_i = \frac{i}{T}$ where $i \in {0,1,2,...,9}$, where $T \approx 10.3$ years is the total time of observations. The higher frequency bins are dominated by white noise, so we do not include them in the analysis; see~\cite{EPTA:2023fyk} for details. We will refer to the EPTA dataset simply as ``EPTA''. 

\subsection{Physical and observational constraints}\label{sec:analys2}

In this work, we implement relevant physical and observational constraints on the IGWB. 
The first group of constraints described below is implemented at the likelihood level, while the second group is implemented \emph{a posteriori}, a form of importance sampling.

\subsubsection{``BLVK'' constraints}

{\bf BBN+CMB bounds.} The integrated GW energy density must not exceed the upper limits set by BBN and CMB observations on additional contributions to the radiation energy density, 
\begin{equation}\label{integral_bounds}
    \int_{f_i}^{\fend}\frac{df}{f}h^2\Omega_{\text{GW}}(f) \leq \mathcal{B}_i\,,\quad i=\{{\rm BBN}, {\rm CMB}\} \, .
\end{equation}
For the lower integration bound, $f_i=\{f_{\text{BBN}},f_{\text{CMB}}\}$, we adopt respectively: the mode that are sub-Hubble at the time of BBN,
$f_{\text{BBN}} = 3\times10^{-13}\,\Hz$ and $f_{\text{CMB}}$ as defined in Sec.~\ref{sec:model}. 
The upper integration bound is set by the maximum frequency associated with the end of inflation, $\fend$.
The BBN bound is parametrized in terms of the allowed number of extra relativistic degrees of freedom at the BBN epoch, $\Delta N_{\rm eff}$, leading to the bound $\mathcal{B_{\text{BBN}}} = 5.6\times10^{-6} \, \Delta N_{\rm eff}$.
We adopt $\Delta N_{\rm eff} = 0.39$ from \cite{Caprini:2018mtu}. 
For CMB, the upper limit is instead $\mathcal{B_{\text{CMB}}} = 6.9\times10^{-6}$ \cite{Caprini:2018mtu}. 
It should be noted that more recent measurements have further strengthened these bounds (even down to $N_{\rm eff}<3$, see e.g.~\cite{Elbers:2025vlz,AtacamaCosmologyTelescope:2025nti,Cielo:2023bqp,Schoneberg:2024ifp}), but we adopt these older values to remain conservative.

For a simple blue-tilted spectrum, the BBN constraint is  more restrictive than the CMB one. However, for a non-trivial IGWB spectrum, the integrand in \eqref{integral_bounds} may receive a large contribution from the frequency interval $f_{\rm CMB}<f<f_{\rm BBN}$, in which case the CMB bound would be more relevant. While this effect is not significant for the constant-tilt spectrum adopted in this work, we still implement both the BBN and CMB constraints in our analysis.

{\bf LVK bound.} We implement the upper limit on an isotropic GWB obtained by the LIGO–Virgo–KAGRA (LVK) collaboration~\cite{LIGOScientific:2025bgj}. 
Specifically, we set 
\be\label{eq:cond2}
 \Omega_{\rm GW} \leq 2.8 \times 10^{-9} \equiv \Omega_{\text{LVK}} \quad \text{at}\quad f_{\rm LVK}\simeq 25\,\Hz \, .
\ee
This bound was derived at a $95\%$ confidence level (CL) assuming a scale-invariant GWB spectrum. 
While the GWB we adopt in this work in general deviates from scale-invariance, the LVK band is narrow with respect to the frequency band under consideration here, and adopting the limit for a scale invariant spectrum does not affect our conclusions.

{\bf Constraint on the cutoff frequency.} We require that the frequency crossing the Hubble scale at the end of inflation is larger than the one crossing the Hubble scale at the end of reheating, 
\be\label{eq:cond3}
\frh \leq \fend\,.
\ee
The limiting case $\frh = \fend$ corresponds to instantaneous reheating.
This condition reflects the monotonic growth of the comoving Hubble radius after inflation during the standard cosmological evolution. While it is expected to hold, it can be violated in certain non-standard cosmological scenarios, such as thermal inflation~\cite{Lyth:1995ka} and bouncing cosmologies~\cite{Brandenberger:2016vhg}.

All three physical constraints described above are implemented at the likelihood level by rejecting MCMC proposals that do not satisfy Eqs.~\eqref{integral_bounds},~\eqref{eq:cond2}~\eqref{eq:cond3}. In what follows, we collectively refer to this set of constraints as ``BLVK''. 

\subsubsection{``CMB'' constraints}

{\bf CMB data.}
The CMB polarization data provide additional information about the amplitude and the tilt of the primordial tensor power spectrum. 
We take into account the measurement of the CMB B-mode polarization angular power spectrum from the joint BICEP2/Keck Array and Planck analysis~\cite{Planck:2018jri} that is, at $95\%$ CL, 
\be\label{eq:cmb}
\begin{split}
    r_{0.01}&<0.076\\
    -0.55 &< n_{t} < 2.54 
\end{split}
\ee
where $r_{0.01}$ denotes the tensor-to-scalar ratio evaluated at the scale $k=0.01\Mpc^{-1}$. These constraints were obtained from measurements of the tensor-to-scalar ratio at two different scales, $k=0.002\Mpc^{-1}$ and $k=0.02\Mpc^{-1}$, assuming a power-law tensor power spectrum.
As demonstrated in Sec.\ \ref{sec:res}, 
the posterior parameter space selected when implementing the BLVK constraints is almost completely contained into the region given in Eqs.~\eqref{eq:cmb}.
Hence, we implement these constraints {\it a posteriori} by selecting only the MCMC samples that satisfy Eq.~\eqref{eq:cmb}.
Note that, in our analysis, $r$ and $A_s$ are defined at the CMB scale; we rescale accordingly the tensor-to-scalar ratio measured at $k=0.01\Mpc^{-1}$ to the CMB scale with $r=r_{0.01}(f_{\rm CMB}/f_{0.01})^{n_t+1-n_s}$. To perform this conversion, we use $n_s = 0.965$, the Planck best fit value \cite{Planck:2018vyg} of the scalar index.

{\bf The scale of inflation.} As explained in Sec.~\ref{sec:model}, the upper bound on $r_{0.01}$, 
together with the scalar spectrum amplitude $A_s$, set 
an upper bound on $H_{\rm inf}$, given in Eq.~\eqref{eq:Hinf}.
Since $H_{\rm end}\leq H_{\rm inf}$, in the context of the present analysis this must be implemented as a constraint on $\fend$, the parameter we have chosen to sample on (c.f.~discussion in Sec.~\ref{sec:analys3}).

First of all, the upper bound on $\fend $ due to instantaneous reheating, given in Eq.~\eqref{eq:fend_value}, is set as an upper prior on $\fend$ (see Sec.~\ref{sec:analys3} and Table \ref{tab:priors}). 
However, the fact that we also account for non-instantaneous reheating indicates that Eq.~\eqref{eq:fend_formula} should be also set as an adaptive constraint on $\fend$.
We implement this constraint \emph{a posteriori}, 
by eliminating all the MCMC samples for which the values of $\fend$ and $T_{\rm reh}$ lead to a violation of the upper bound on $H_{\rm end}\leq H_{\rm inf}$, inferred from Eq.~\eqref{eq:Hinf}.
This becomes very constraining in particular 
in light of the fact that the prior we adopt on $\Trh$ (see Sec.~\ref{sec:analys3} and Table \ref{tab:priors}) selects temperatures that are very low compared to what is
natural in the context of instantaneous reheating: 
\begin{equation}\label{eq:Treh_max}
    T_{\rm rh}=\sqrt{\frac{H_{\rm end}M_{\rm Pl}}{\pi}}\,\left(\frac{90}{g_{*}^{\rm rh}}\right)^{1/4}\lesssim 7 \times 10^{15}\,{\rm GeV}\,.
\end{equation} 
However, it is justified to implement this constraint \emph{a posteriori} since, as we will see, it only affects an uninformative part of the posterior parameter space.

{\bf The number of e-folds of reheating.} 
Finally, one last physical constraint relevant in our analysis concerns the number of e-folds of reheating, given by
\begin{equation}\label{eq:nrh}
    N_{\rm rh} = \ln \frac{a_{\rm rh}}{a_{\rm end}}= \ln\frac{\frh}{\fend} - \ln\frac{H_{\rm rh}}{H_{\rm end}}\,.
\end{equation}
As already shown in \cite{NANOGrav:2023hvm}, it is important to implement the lower bound $N_{\rm rh} > 0$ in the analysis. 
We also do this \emph{a posteriori} on the MCMC samples, as the region of parameter space in which it plays a major role is already disfavoured by the BBN and CMB bounds. 

MCMC samples that are selected according to Eq.~\eqref{eq:cmb}, $H_{\rm end}\leq H_{\rm inf}$, and $N_{\rm rh}>0$ are referred to as ``CMB'' in what follows.

\subsection{Analysis procedure}\label{sec:analys3}

{\bf EPTA data analysis.} The analysis of the EPTA data is performed using the \texttt{ENTERPRISE} software package~\cite{Ellis:2020enterprise}. The EPTA analysis pipeline is described in detail in \cite{EPTA:2023fyk}; here we summarize its main elements.

We fix the white noise parameters of each pulsar to the maximum likelihood values obtained from the single pulsar noise analysis~\cite{EPTA:2023akd}. We adopt a simplified noise model including red noise (RN) and dispersion-measure variations (DM) only for every pulsar.
The RN and DM contributions are described by power law spectra specified by four free parameters per pulsar $\theta_{{\rm PTA}}=\{\log_{10}A_{\text{RN}}, \gamma_{\text{RN}},\log_{10}A_{\text{DM}}, \gamma_{\text{DM}}\}_{i}$, sampled using priors from \cite{EPTA:2023fyk}. It should be noted that the pulsar J1713+0747 also manifests one exponential dip in DR2New. Hence, the additional chromatic noise parameters describing the amplitude $A$, relaxation time $\tau$ and epoch $t_0$ are also modelled for this particular event.  More information on the priors and model used for this dip can be found in~\cite{EPTA:2023akd}.  

In our analysis, the common red signal in all pulsars is modeled within the Common Uncorrelated Red Noise (CURN) framework. This is done 
since it significantly accelerates the analysis by reducing the complexity of the MCMC sampling. 
Considering the large uncertainties in the parameters of the observed signal, we do not expect the inclusion of spatial correlations to substantially affect our results.
We implement the effects of SSE systematics using a fixed SSE model, DE440, from~\cite{Park_2021}, following the official collaboration analysis.

{\bf Cosmological parameters.} Along with the PTA parameters, we sample the cosmological parameters of the theoretical model (see Sec.~\ref{sec:model}). 
Tab.~\ref{tab:priors} summarizes the cosmological parameters and their adopted priors.

For $n_t$, $r$ and $\Trh$, we adopt similar priors to those used in the official NANOGrav analysis~\cite{NANOGrav:2023hvm}. 
This choice facilitates the comparison with the results of the NANOGrav collaboration presented in Sec.~\ref{sec:comparison}.
The prior on $\Trh$ is set to values around the PTA frequency band: a frequency of the order of $f_{\rm PTA}\simeq 10^{-8}$ Hz corresponds indeed to $T_{\rm PTA}\simeq 0.5\,{\rm GeV}$, as can be derived from Eq.~\eqref{eq:frh} applied to a generic $T$ in the radiation era, and setting $g_{*,s}=g_*\simeq 10$ as appropriate around the GeV scale. 
$T_{\rm PTA}$ is clearly much lower than the current upper bound that can be inferred on the reheating temperature from the energy scale of inflation if one assumes instantaneous reheating, Eq.~\eqref{eq:Treh_max}.
Further discussion on the consequences of setting this prior to such low values will be made in Sec.~\ref{sec:res}.

\begin{table}[t]
    \centering
    \begin{tabular}{l@{\hspace{2em}}l}
    \hline\hline
    Parameter & Prior\\
    \hline
        $\log_{10}r$  &  $[-40, 0]$  \\ 
        $n_t$  &  $[-1, 6]$ \\
        $\log_{10}(\Trh/\GeV)$   &  $[-3, 3]$ \\
        $\log_{10}(\fend/\Hz)$   &  $[-9, 8.2]$ \\
    \hline
    \end{tabular}
    \caption{Uniform priors for the cosmological parameters which are directly sampled in the MCMC analyses.
    }
    \label{tab:priors}
\end{table}

Concerning the prior on the inflationary cut-off of the IGW spectrum, $\fend$, the lower bound is $f_{\rm end} = 10^{-9}\,\Hz$, corresponding to the lowest frequency in the PTA frequency range, ensuring a inflationary interpretation of the EPTA signal.
The upper bound is $f_{\rm end}= 1.5 \times 10^8$ Hz, i.e.~the value of $f_{\rm end}$ derived under the assumption of instantaneous reheating from the measurement of CMB anisotropies, Eq.~\eqref{eq:fend_value}.
This is well above the LVK frequency scale~\cite{LIGOScientific:2025bgj}, ensuring that the physically relevant region of $\fend$ is adequately sampled in the MCMC analysis.

Note that $\fend$ is related to $\Trh$ though Eq.~\eqref{eq:fend_formula}. We could have therefore chosen as a parameter of the model $H_{\rm end}$ instead on $\fend$. 
We chose to sample in terms of $\fend$ as its interpretation is more intuitive: it directly corresponds to the upper cutoff of the IGWB spectrum,  making its role clearer than that of $H_{\rm end}$. 
In Sec.~\ref{sec:derived_parameters} we discuss derived parameters, including $H_{\rm end}$.

{\bf Statistical analysis.} We perform a MCMC analysis to sample from 
the posterior distribution using the Metropolis-Hastings algorithm with Parallel Tempering~\cite{ellis_jellis18ptmcmcsampler_2017_2}, as implemented in the \texttt{ENTERPRISE} code.
This technique runs multiple MCMC processes at different
temperatures, allowing the swap of information between them and, thereby, enabling a more efficient exploration of the multi-dimensional parameter space. Parallel Tempering is especially effective for sampling multi-peaked posterior distributions, which is the case in the EPTA data analysis when no additional observational constraints are imposed. Throughout this analysis, 8 parallel chains are used, set to temperatures following the relation:
\begin{equation}
    T_i = T_0\left(1 + \sqrt{\frac{2}{N}}\right)^i
\end{equation}
where $T_0 = 1$, $N = N_{\text{RN}} + N_{\text{DM}} +N_{\text{dip}} + N_{\text{CURN}} = 107$ is the number of parameters sampled and $i$ is the label of the parallel chain. This provides an efficient exploration of the parameter space and enables adequate sampling of multiple posterior modes, that are expected to manifest due to the different regimes of horizon re-entry allowed for the IGWB.

The plots and marginalized constraints are computed using the public \texttt{corner} code~\cite{ForemanMackey:2016corner}. We adopt a Gelman-Rubin convergence criterion $|R - 1| < 0.1$ for the main analyses of this work, which is calculated with \texttt{getDist} \cite{Lewis:2019xzd}. Finally, the median and confidence intervals are evaluated with \texttt{numpy} \cite{harris2020array}.
\section{Results}\label{sec:res}

\begin{figure*}
    \centering\includegraphics[width=1.3\columnwidth]{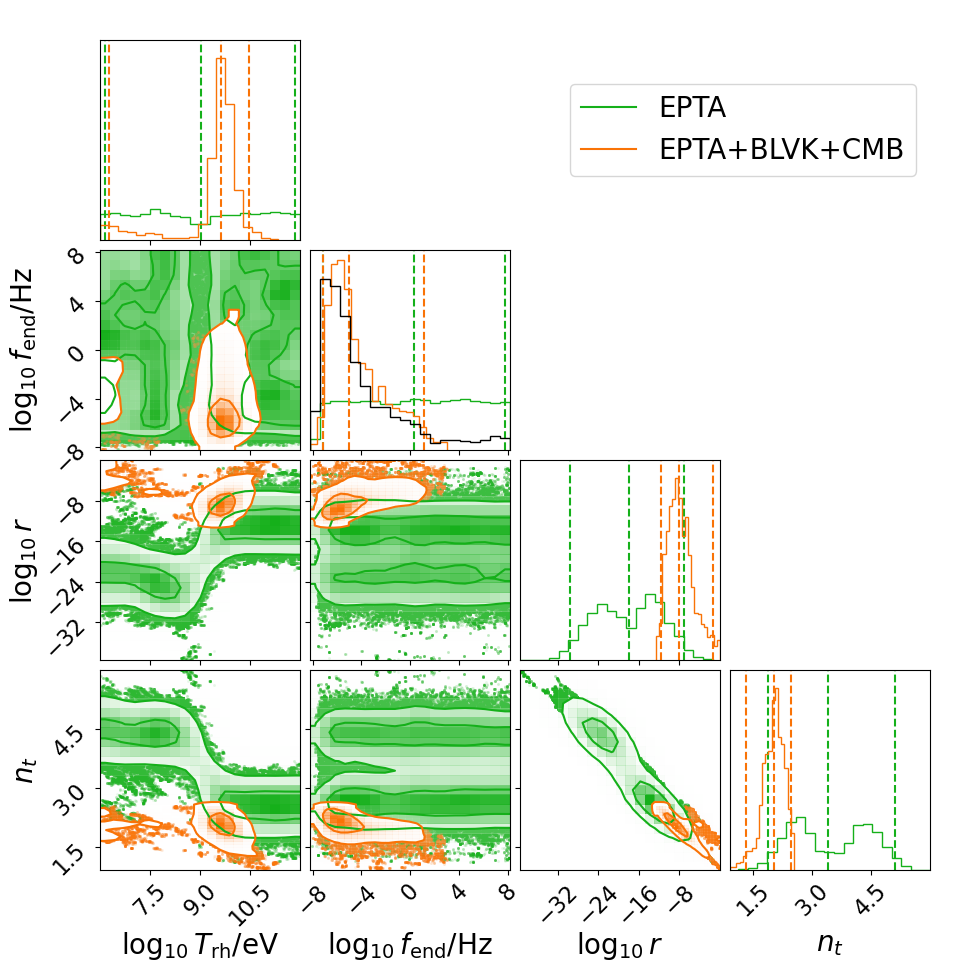}
    \caption{
    Posterior distributions for the $\Trh$, $\fend$, $r$ and $n_t$ parameters obtained from the EPTA (green) and EPTA+BLVK+CMB (orange) datasets. The contours smooth the data in the $1\sigma$ and $2 \sigma$ credible regions. The posterior distributions obtained in the EPTA+BLVK analysis aren't represented in this figure, since they only differ from the
    EPTA+BLVK+CMB posteriors by a 
    small
    portion of parameter space. The effect of imposing the parameter space cut corresponding to Eqs.~\eqref{eq:cmb} \emph{a posteriori} is visible in terms of the upper 
    flat cut in the $(\Trh-n_t)$ posterior, and the lower flat cut in the $(\Trh -\log_{10}r)$ one (and correspondingly for $(\fend-n_t)$ and $(\fend-\log_{10}r)$). 
    Imposing the constraints on the scale of inflation $H_{\rm end}\leq H_{\rm inf}$ has the effect of cutting the tail of the posterior distribution on $\fend$, as can be appreciated by comparing the 1D marginalized histogram of $\fend$ of the EPTA+BLVK dataset 
    (in black) with the one of the 
    EPTA+BLVK+CMB dataset 
    (in orange).}
    \label{fig:cosmo}
\end{figure*}

\begin{figure}[!t]
    \includegraphics[width=0.99\columnwidth]{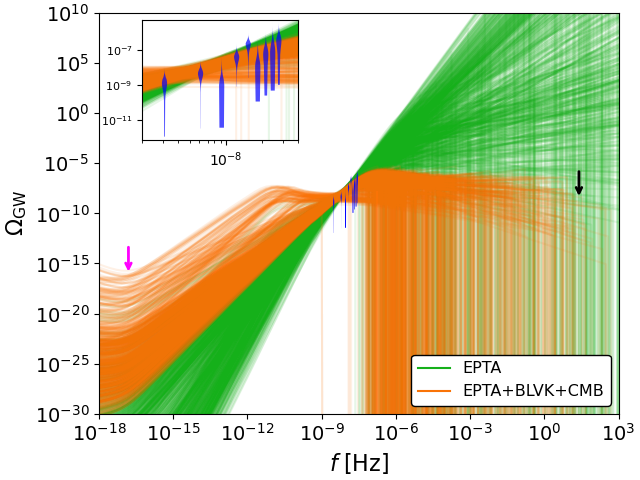}
    \caption{Theoretical predictions for $\Omega_{\rm GW}(f)$ randomly selected from the chains in the EPTA (green) and EPTA+BLVK+CMB (orange) data analyses. The curves correspond to a 1\% subsample of the respective chains. The blue violin plots represent the EPTA measurements obtained with the free spectrum method by sampling nine discrete EPTA frequencies~\cite{EPTA:2023fyk}. The downward black arrow on the right indicates the $95\%$ upper limit on the GWB amplitude reported by the LIGO–Virgo–KAGRA (LVK) collaboration~\cite{LIGOScientific:2025bgj}, while the downward magenta arrow on the left represents the CMB bound on $r$ at the pivot scale $f_{\rm CMB}$~\cite{Planck:2018jri}.}
    \label{fig:OmGW}
\end{figure}

Fig.~\ref{fig:cosmo} shows the main results on the cosmological parameters obtained from the chains.
Tab.~\ref{tab:param1} presents the one-dimensional marginalized constraints.
Fig.~\ref{fig:OmGW} displays the theoretical predictions of $\Omega_{\rm GW}(f)$ for a subset of chains randomly selected from these analyses. 

\begin{table*}[!t]
    \centering
    \begin{tabular}{lcccc}
    \hline\hline
    Data/Model 
    & $\log_{10}r$
    & $n_t$
    & $\enspace \log_{10}(\Trh/\eV)\enspace$
    & $\enspace \log_{10}(\fend/\Hz)\enspace$
    \\\hline
    EPTA
    & $\enspace -18.02^{+10.96}_{-11.67}\enspace$
    & $\enspace 3.42^{+1.70}_{-1.53}\enspace$
    & Unconstrained
    & $ > -7.16$
    \\
    EPTA+BLVK
    & $-8.23^{+6.87}_{-5.09}\enspace$
    & $2.05^{+0.77}_{-0.76}\enspace$
    & $9.65^{+0.83}_{-3.39}\enspace$
    & $-5.25^{+11.98}_{-2.24}\enspace$
    \\
    EPTA+BLVK+CMB
    & $-8.11^{+6.66}_{-3.55}\enspace$
    & $2.04^{+0.43}_{-0.72}\enspace$
    & $9.64^{+0.81}_{-3.39}\enspace$
    & $-4.98^{+6.14}_{-2.14}\enspace$
    \\\hline
    \end{tabular}
    \caption{Mean and 95\% CL or lower bounds for the cosmological parameters in the EPTA, EPTA+BLVK and EPTA+BLVK+CMB analyses. The two-dimensional posteriors are shown in Fig.~\ref{fig:cosmo}.}
    \label{tab:param1}
\end{table*}

\subsection{EPTA analysis}\label{sec:res1}

We begin by presenting results of the EPTA data analysis obtained without imposing external constraints. 

In the EPTA-only analysis, the posterior distributions of $r$, $n_t$ and $\Trh$ exhibit a clear bimodal structure. This is consistent with the findings of the NANOGrav 15-years dataset analysis \cite{NANOGrav:2023hvm}, see Sec.~\ref{sec:comparison}. 
In the IGW model, one can identify two different regimes: $\Trh> 1\,\GeV$ and $\Trh< 1\,\GeV$. In the first case, $\fpta< \frh$ according to Eq.~\eqref{eq:frh}, taking $\fpta\simeq10^{-8}$ Hz, and the IGWB spectrum in the PTA frequency band is generated by tensor modes that re-entered the horizon during the radiation-dominated era. 
For $\Trh < 1\,\GeV$, on the other hand, one has $\frh< \fpta$, and the GW spectrum in the PTA frequency band is generated by tensor modes that re-entered the horizon during an extended reheating phase occurring at very low energy. 
Note that the posterior distribution of $\Trh$ is dominated by the prior.
Therefore, the EPTA data alone do not constrain the reheating temperature. Even if our prior would extend to values closer to the maximal expected reheating scale, see Eq.~\eqref{eq:Treh_max}, no value would be preferred and the posterior would stay flat up to these large values.  

On the other hand, the EPTA data provide informative constraints on the spectral tilt of the common red signal that, in our case, transform into a measurement of $n_t$.
The best-fit EPTA value of the spectral index of the timing residual power spectral density $S(f)$ given in Eq.~\eqref{eq:S} is $\gamma\simeq 2.9$ in the DR2New+CURN analysis with \texttt{Enterprise} \cite{EPTA:2023fyk}.
Assuming a cosmological origin of the PTA signal, this quantity is related to the slope of the $\Omega_{\text{GW}}(f)$ spectrum at $f \simeq \fpta$. 
When $\frh > \fpta$, the tilt of the IGW spectrum at PTA frequencies is directly given by $n_t$, since the transfer function is $\propto f^2$ (c.f.~Eq.~\eqref{eq:OmGW} and discussion in Sec.~\ref{sec:model}). 
The equivalent in terms of $\gamma$ would be $\gamma = 5 - n_t$ \cite{EPTA:2023xxk}. 
When $\frh < \fpta$, the tilt of the IGW spectrum is instead $n_t-2$, since the transfer function is $\propto f^0$, hence an equivalent of $\gamma = 7 - n_t$. As a result, the data select different values of $n_t$ in the two regimes.

In particular, in the large-$\Trh$ regime (modes entering the horizon during the radiation dominated era), the data favour $n_t\approx 2$, while in the small-$\Trh$ case (modes entering the horizon during reheating) they prefer $n_t\approx 4$; in both cases, $\gamma\approx 3$ is reproduced at PTA frequencies in full agreement with~\cite{EPTA:2023fyk}.
In addition, there is a strong correlation between the spectral index $n_t$ and the tensor-to-scalar ratio $r$, arising from the fact that these parameters enter in the combination $ \Omega_{\rm GW}(f_{\rm PTA})\propto r\, \left( f_{\rm PTA}/f_{\rm CMB} \right)^{n_t}$ (see~Eq.~\eqref{eq:OmGW}), with $f_{\rm CMB}\ll f_{\rm PTA}$.  
This degeneracy leads to different preferred values of $r$: larger values of $r$ for $\Trh> 1\,\GeV$ 
when the best fit IGWB spectrum is shallower $n_t\approx 2$, 
and smaller values of $r$ for $\Trh< 1\,\GeV$, when the best fit IGWB spectrum is steeper $n_t\approx 4$.
Overall, the two regimes are characterized by
\begin{align}\label{eq:regL}
&\Trh< 1\,\GeV:\qquad (n_t,\log_{10}r)\sim(4,-26)\\
\label{eq:regH}
&\Trh> 1\,\GeV:\qquad (n_t,\log_{10}r)\sim(2,-15)
\end{align}
Since the PTA data prefer a positive tilt, the corresponding values of $r$ at the CMB scales are extremely low compared to the current upper bounds from CMB (see Eq.~\eqref{eq:cmb}) and to the typical values expected within single-field slow roll scenarios. 
Furthermore, because of the relatively low constraining power of the current PTA data, combined with the large frequency span between PTA and CMB scales, 
the posterior of $r$ is also extremely wide, spanning more than 20 orders of magnitude. 
Nonetheless, among the two regimes in Eqs.~\eqref{eq:regL}, \eqref{eq:regH} identified by the PTA data analysis, the one characterised by low $n_t$ and higher $r$ is more natural. 
As we will see in the next section, this region of the parameter space is the one selected when accounting for the observational constraints.

The remaining cosmological parameter, $\fend$, shows an approximately flat posterior distribution. 
The lower bound on $\fend$ is determined by the PTA frequency band, which is required to fit 
the EPTA signal; while the upper bound is set by the prior. 
We conclude that, in the absence of additional physical constraints, the EPTA data are not sensitive to the high-frequency cutoff of the IGWB. 

\subsection{EPTA+BLVK+CMB analysis}\label{sec:res2}

In this section, we discuss the implications of incorporating the BLVK and CMB constraints.

Fig.~\ref{fig:cosmo} and Tab.~\ref{tab:param1} clearly show that including the BLVK constraints has a profound impact on the EPTA data analysis.
In particular, it breaks the degeneracy between the ``large''-$\Trh$ and ``small''-$\Trh$ regions of parameter space (large and small within the prior of $\Trh$ considered). 
As previously mentioned, the EPTA+BLVK data favour the more natural region of parameter space given in Eq.~\eqref{eq:regH}, featuring larger reheating temperatures,\footnote{Although the region associated with the smaller reheating temperatures is disfavoured by the data, it is not fully excluded, leading to a small tail in the posterior distribution at low $\Trh$.} lower $n_t$ and larger $r$. 
This occurs because larger $\Trh$ values imply $\frh>\fpta$, and therefore the EPTA data are fit by modes entering the horizon in the radiation dominated era, implying smaller values of the tensor spectral index $n_t\approx 2$. 
Correspondingly, a flatter GWB spectrum features less power in the LVK frequency band, and a smaller integrated GW energy density, suppressed enough to obey the BBN+CMB integrated energy density constraint. 
On the other hand, the small-$\Trh$ region typically requires significantly larger values of $n_t\approx 4$, 
violating the LVK and the BBN+CMB bounds.

An important implication of our results is that including the BLVK bounds in the EPTA data analysis leads to a prediction for $\Trh$. 
This occurs in two ways. First, as discussed above, the ``small''-$\Trh$ region of parameter space is disfavoured. Second, although the data favour the ``large''-$\Trh$ regime, corresponding to $\Trh > 1\,\GeV$, we find that the reheating temperature cannot be significantly higher than the value associated with PTA frequencies, i.e.~$\fpta \lesssim \frh$. Indeed, large values of $\Trh$ enhance the IGWB spectrum integrated energy density, since $\Omega_{\rm GW}(f)\propto f^{n_t}$ at $f<\frh$ with $n_t>0$, violating the BBN+CMB bound. As a result, the data select a specific window, $1.78  \,\rm{MeV} \lesssim \Trh \lesssim 28.2 \,\rm{GeV}$, where the upper bound is set by the BBN+CMB constraint. Such low values of $\Trh$ are very challenging to explain theoretically.

As evidenced in Tab.~\ref{tab:param1}, the parameter constraints obtained from the EPTA+BLVK and EPTA+BLVK+CMB analyses are very similar for $r$, $n_t$ and $\Trh$, because imposing the BLVK bounds selects a region of parameter space that practically satisfies Eqs.~\eqref{eq:cmb}. 
When imposing the CMB polarization data constraints \emph{a posteriori}, a small portion of the $n_t$ parameter space with too large values is excluded; this in turn implies that the corresponding low values of $r$ are also excluded (c.f.~Fig.~\ref{fig:cosmo}).  

On the other hand, the posterior distribution of $\fend$ is appreciably altered by including CMB constraints as opposed to only including BLVK, as can be appreciated from the confidence intervals listed in Table \ref{tab:param1}.

It can be seen that including the BLVK constraints changes the posterior from flat (in the EPTA-only analysis, see Sec.~\ref{sec:res1})
to peaking at frequencies just above the PTA band, $\log_{10}{\fend}\gtrsim -7.5$. 
Despite the relatively shallow blue-tilted IGW spectrum with $n_t\lesssim 3$, 
the frequency range $\frh<f<\fend$ where $\Omega_{\rm GW}(f)\propto f^{n_t-2}$ produces a substantial contribution to the integrated GW energy density, which is constrained by the BBN+CMB bound.
Introducing an early cutoff of the IGWB spectrum at $\fend\gtrsim\frh$ effectively alleviates the BLVK constraint. 

However, for values of the tensor tilt $n_t\lesssim 2$, for which $n_t-2$ is negative and the integral \eqref{integral_bounds} converges, the allowed range of the cutoff frequency can extend to higher values, which can be formally separated into two cases: $\fend<\flvk$ and $\fend>\flvk$.
In the former case, the $\Omega_{\text{GW}}$ spectrum is truncated below the LVK frequencies, $\flvk \simeq 25\,\Hz$, effectively avoiding the LVK spectral energy density bound. 
In the latter case, the BLVK constraints remain compatible with an arbitrary duration of reheating, and the posterior of $\fend$ has a tail that is cut off only by the prior: see a $95 \%$ CL given in Table.~\ref{tab:param1} and the $\fend$ distribution in Fig.~\ref{fig:cosmo} plotted in black, representing the EPTA+BLVK dataset.
This region of the parameter space with large $\fend$ is the one for which the $H_{\rm end}\leq H_{\rm inf}$ constraint plays a relevant role. 
An arbitrary duration of reheating is incompatible with this constraint, which 
effectively cuts off the BLVK posterior tail with $\fend\gtrsim 2 \cdot 10^3$ Hz (this value can be derived from Eq.~\eqref{eq:fend_formula} setting $H_{\rm end}= H_{\rm inf}$ and $\Trh$ to its upper 95\% confidence value).

In the EPTA-only analysis, the posteriors are highly non-Gaussian, which implies that there are projection effects. 
Such effects arise from the marginalization over parameters in the presence of non-Gaussian posterior correlations, which shift the marginalized posterior away from the maximum a posteriori value.
According to Fig.~\ref{fig:cosmo}, the $2\sigma$ posteriors of $n_t$ and $r$ associated with the region of parameter space where $\Trh > 1$ GeV are shifted toward larger and smaller values, respectively, compared to those in the EPTA+BLVK+CMB analysis. These shifts can be partially attributed to unconstrained parameter directions in the EPTA-only analysis, which enhance projection effects. The inclusion of BLVK+CMB data provides informative constraints on the model parameters, reducing these effects in the joint analysis.

\subsection{Inferred duration of reheating and the scale of inflation}
\label{sec:derived_parameters}

In this analysis, we have chosen to sample directly on  $\Trh$ and $\fend$. 
However, these parameters are related to two relevant physical parameters, namely $N_{\rm rh}$ and $H_{\rm end}$. In this section, we investigate what can be learned from our posteriors on these two derived parameters.

\begin{figure}[!h]
    \centering
\includegraphics[width=0.99\columnwidth]{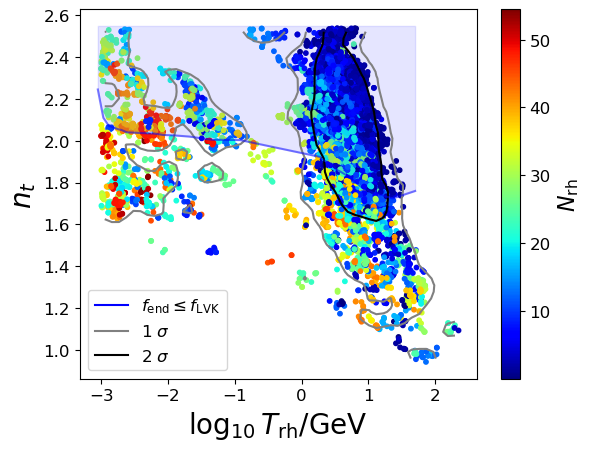} 
    \caption{Number of e-folds of reheating in the posterior parameter space $\log_{10}\Trh$ and $n_t$ for the EPTA+BLVK+CMB dataset. 
    The light-blue shaded region corresponds to the portion of parameter space that would lead to
    $\Omega_{\text{GW}}( \flvk) \geq \Omega_{\text{LVK}}$ (see Eq.~\eqref{eq:cond2}).}
    \label{fig:N_rh}
\end{figure}

We start by deriving the distribution on the duration of reheating. 
For a given sample featuring $\Trh$ and $\fend$, the number of e-folds of reheating, given by Eq.\ \eqref{eq:nrh}, 
becomes
\begin{equation}
    N_{\rm rh}=\ln \left[\frac{360}{g_*^{\rm rh}}\left(\frac{g_{*,s}^{\rm reh}}{g_{*,s}^{0}}\right)^{2/3}  \left(\frac{M_{\rm Pl}}{T_0}\right)^2
    \left(\frac{\fend}{\Trh}\right)^2\right]\,,
\end{equation}
obtained using Eqs.~\eqref{eq:frh}, \eqref{eq:fend_formula} and \eqref{eq:Treh_max} with $H_{\rm end}$.
The result is shown in Fig.~\ref{fig:N_rh} for the EPTA+BLVK+CMB data.
The samples with constrained reheating length, $N_\text{rh} \lesssim 10$, are about $\sim 60\%$ of the samples and are mainly associated with a large spectral index $n_t \gtrsim 1.8$ and $\Trh>1$ GeV. 
These correspond to modes that entered the horizon during the radiation dominated era, $f_{\rm rh}>f_{\rm PTA}$, have a blue spectrum in the frequency range $f_{\rm rh}< f<\fend$, and are therefore cut-off
by the LVK bound of Eq.~\eqref{eq:cond2}, so that $\fend \leq \flvk$ (see the light-blue shaded area in Fig.~\ref{fig:N_rh}). 
For the rest of the samples, the duration of reheating is controlled by the other BLVK+CMB constraints.
This implies that reheating models following inflationary models that are fitting PTA data must have appropriately tuned duration. 

\begin{figure}
    \centering
\includegraphics[width=0.99\columnwidth]{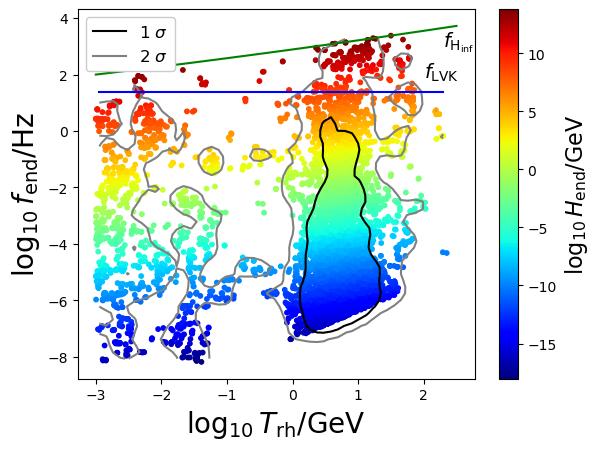}
    \caption{Hubble factor at the end of inflation in the posterior parameter space $\log_{10}\Trh$ and $\log_{10}\fend$ for the EPTA+BLVK+CMB dataset. 
    The power-law relation between $\fend$ and $\Trh$ of Eq.~\eqref{eq:fend_formula} can be appreciated in the colour pattern. 
    The green line shows the constant-$H_{\rm end}$ cut-off induced by the constraint $H_{\rm end}\leq H_{\rm inf}$. Samples with high values of $H_{\rm end}$ are those with $\fend\geq f_{\rm LVK}$, shown by the blue horizontal line.
    Note that the Gaussian density estimate of the $1\sigma, 2\sigma$ confidence intervals (black and grey lines respectively) includes a portion of the parameter space that is in reality excluded by the $\frh \leq \fend$ bound (bottom-right part of the plot).}
    \label{fig:H_end}
\end{figure}

Another relevant physical quantity of an inflationary model 
is the Hubble factor at the end of inflation $H_{\rm end}$, related to $\fend$ and $\Trh$ through Eq.~\eqref{eq:fend_formula}.
Fig.~\ref{fig:H_end} shows that most of the samples fitting the EPTA+BLVK+CMB data
have very small Hubble factor at the end of inflation, with the $1 \sigma$ confidence interval satisfying $1 \rm \, \mu eV \lesssim H_{\rm end} \lesssim 1 \rm \, GeV$. These low values are connected to the correspondingly low values of $\Trh$ and $\fend$ necessary to fit PTA data with inflation, and are very challenging to account for theoretically.

\section{Comparison to previous analyses}\label{sec:comparison}

\begin{figure}[!h]
    \centering
    \includegraphics[width=0.8\linewidth]{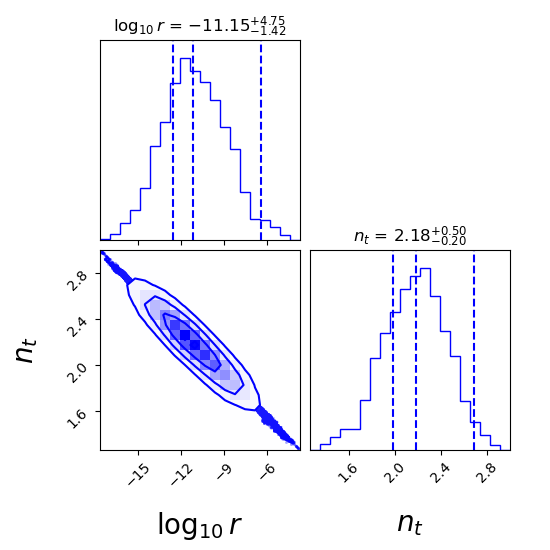}
    \caption{95\% CL posteriors for the parameters $n_t$ and $\log_{10}r$ for the model given in Eq.~\eqref{Single_Pivot_Omega}.}
    \label{fig:single}
\end{figure}

In this section we compare our results with results derived in previous analyses, starting with
the EPTA official DR2 analysis \cite{EPTA:2023xxk}. 
In this work, they adopt an inflationary model with a single spectral pivot \cite{Lasky:2015lej},
\begin{eqnarray}
    \lefteqn{\Omega_{\text{GW}}(f) =} \nonumber\\ 
    &\frac{3}{128}rA_{\text{s}}\Omega_{\text{rad}}\left(\frac{f}{f_{\text{CMB}}}\right)^{n_t}\left[\frac{1}{2}\left(\frac{f_{\text{eq}}}{f}\right)^{2} + \frac{16}{9}\right]\,,
    \label{Single_Pivot_Omega}
\end{eqnarray}
considered in the limit $f\gg f_{\rm eq}$. 
The parameter space is $(r-n_t)$, with a prior on $n_t \in [-1,3]$.
As a consistency check, we have performed an analysis using the model in Eq.~\eqref{Single_Pivot_Omega}, but adopting the priors used throughout this work (see Tab.~\ref{tab:priors}). 
The posterior distribution is presented in Fig.~\ref{fig:single}, while  Tab.~\ref{tab:comparison_single} compares the results obtained in~\cite{EPTA:2023xxk} to our results.  
They are consistent, although the $95 \%$ CL obtained in~\cite{EPTA:2023xxk} is larger than ours. 
This can be explained through three considerations: a) while we sample our parameters through \texttt{ENTERPRISE}'s full noise analysis, EPTA applies Bayesian inference on data obtained with the free spectrum method to calculate likelihoods, as elaborated in \cite{EuropeanPulsarTimingArray:2023lqe}. 
While faster than the method used throughout this work, their method sacrifices some accuracy when calculating the joint distributions of strains across different frequency bins, as these assume that each frequency bin is statistically independent. b) EPTA implements HD correlations into their pipeline, which are not considered in our analysis. 
c) EPTA uses a detailed noise model for each pulsar containing DM, RN and scattering variations, and only propagates the relevant noise parameters in the GW analysis  \cite{EPTA:2023akd},
while in this work we always fit both DM and RN to our noise model for each pulsar, as explained in Sec.~\ref{sec:analys3}. 
As a result, some of the IGWB signal may be absorbed through excessive corrections, mainly due to DM variations. 

\begin{table}[!t]
    \centering
    \begin{tabular}{l@{\hspace{2em}}l@{\hspace{2em}}l}
    \hline\hline
     Parameters & $\log_{10}r$ & $n_t$ \\
     \hline
    EPTA DR2 \cite{EPTA:2023xxk}  &  $-12.18^{+8.81}_{-7.00}$ & $2.29^{+0.87}_{-1.11}$ \\
    Single Pivot model & $-11.15^{+4.75}_{-1.42}$ & $2.18^{+0.50}_{-0.20}$ \\
    \hline
    \end{tabular}
    \caption{Mean and 95\% CL on the parameter set ($n_t,r$) of model \eqref{Single_Pivot_Omega}, obtained by the EPTA DR2 analysis \cite{EPTA:2023xxk} and in this work.
    }
    \label{tab:comparison_single}
\end{table}

We also compare our results with the official analysis of the 15-year NANOGrav dataset~\cite{NANOGrav:2023hvm}. 
In this work, they adopt the transfer function given in Eqs.~\eqref{eq:tran}-\eqref{eq:T2} but, as in the EPTA DR2 analysis, they only consider frequencies $f\gg f_{\rm eq}$, so $T_1^2(f)\rightarrow 3.42 (f/f_{\rm eq})^2$ (c.f.~Eq.~\eqref{eq:T1}). 
Furthermore, they only sample three of the model parameters: $(n_t,r,\Trh)$. 
Instead of sampling $\fend$, they infer it by 
inverting Eq.~\eqref{eq:fend_formula} and setting the Hubble factor at the end of inflation to a maximally allowed value compatible with the BBN and LVK bounds.
They also use a slightly weaker BBN bound than us, $\Delta N_{\text{eff}} = 0.5$, and a slightly higher LVK upper bound ~$ \Omega_{\rm GW} \leq 1.7 \times 10^{-8}$.
These bounds are then expressed in reference to the values of $N_{\rm rh}$ that they allow, by shading the parameter space area that violates them. 

After taking into account the BBN and LVK constraints, \cite{NANOGrav:2023hvm} argues that the most likely region of parameter space capable of fitting their  measurements is the one with $\Trh < 1$ GeV, high values of $n_t\approx 4$ and small values of $r$. 
This contrasts with the results of our EPTA+BLVK analysis, which instead excludes this region and favours the one with $\Trh > 1$ GeV, smaller values of $n_t\approx 2$, larger values of $r$, and moderated $N_{\rm rh} \lesssim 20$ e-folds. 
However, the complete posteriors presented in \cite{NANOGrav:2023hvm} are still compatible with the region of parameter space favoured by our analysis.
The difference in the outcome of the two analyses can be attributed to the implementation of the observational constraints. 
The NANOGrav collaboration accounted for the BLVK constraints after processing the MCMC chains, whereas we incorporate them directly at the likelihood level, or \textit{a priori}. 
This allows us to consistently propagate the effects of the observational constraints into the cosmological parameter constraints, leading to more accurate results.

Note that in our analysis, some weight of the posterior is also given to the region with $\Trh \approx 10$ MeV, $n_t \approx 2$, large values of $r$ and more freedom in the duration of reheating. This peak can be seen in Fig. \ref{fig:OmGW} as the curves that have their second pivot around $f \sim 10^{-12} \, \rm Hz$ and that more closely saturate the CMB bound on $r_{0.01}$.

\section{Conclusion}\label{sec:conc}

In this work, we have investigated the parameter space within which an IGWB could account for the EPTA DR2 measurement. 
We have demonstrated that it is important to include external observational and physical constraints in the parameter space selection. This shifts the parameter space of the IGWB capable of fitting PTA data towards more natural values, which remain, however, incompatible with slow roll expectations and extremely challenging to explain even within less standard inflationary models. 

We started by determining the parameter space that would fit the EPTA dataset alone, finding consistent results with the analysis of NANOGrav \cite{NANOGrav:2023hvm}, but extending the latter with one extra parameter, the cutoff frequency representing the end of inflation. We confirm that the parameter space is highly degenerate in the absence of more information. 
Therefore, it is essential to apply constraints from pre-existing measurements. We have considered a set of constraints including the BBN+CMB bound, the LVK bound, CMB measurements and the scale of inflation.  
The results obtained by applying these constraints break the degeneracy of the EPTA dataset, favouring (at 95\% CL) an inflationary scenario with a blue-tilted power spectrum $1.32 \lesssim n_t \lesssim  2.47$, clearly outside the slow roll scenario, and a tensor-to-scalar ratio $r \in [2.2 \cdot 10^{-12},0.035]$ that extends from values well below to values that near the upper bound set by the CMB.  
Interestingly, the EPTA+BLVK+CMB dataset also constrains the temperature and duration of reheating, along with the Hubble parameter at the end of inflation. The typical energy scale of the reheating scenarios allowed is around the PTA scale, yielding reheating temperatures  $1.78 \,{\rm MeV} \lesssim \Trh \lesssim 28.2 \, {\rm GeV}$ and favouring a duration of reheating around $N_\text{rh} \lesssim 20$ e-folds with cutoff frequency at the end of inflation $75.86\,{\rm n\Hz} \lesssim \fend \lesssim 14.45\, {\rm\Hz}$. 
This is driven by a combination of factors: the PTA frequency band that is very low with respect to the expected inflationary scale, the positive spectral index necessary to explain the PTA data, and the observational and physical constraints that limit the overall amount of GW power. 

Not surprisingly, our work further supports that inflation in the simple parametrisation we adopted cannot be considered a natural explanation of the PTA measurement. 
Nevertheless, the framework we developed leads to possibilities for future analyses: for instance, extending our analysis to explore alternative inflationary models. 
These could be based on the effective field theory of inflation \cite{Cheung:2007st,Bianchi:2024qyp}, or on theories of modified gravity such as Gauss-Bonnet inflation \cite{Yin:2024ccm,Bernardo:2025lie}, which could accommodate blue-tilted power spectra. 
One could also investigate more sophisticated reheating models following~\cite{Drewes:2014pfa,Garcia:2020wiy,Ahmed:2021fvt}. 
Most importantly, we have demonstrated that including relevant observational and physical constraints \emph{a priori} in the analysis is important since it substantially changes the selected parameter space.

\section{Acknowledgements}
PT thanks his PhD supervisor Bence B\'ecsy, Golam Shaifullah who wrote the tutorials that served as a foundation for analysing the dataset, and Hippolyte Quelquejay Lecl\`ere who provided information and references on EPTA's methods. CC and PT thank Stas Babak for advice on implementing certain functions in \texttt{ENTERPRISE}. The authors also thank Amodio Carleo for comments on the manuscript. DP acknowledges financial support from the INAF Large Grant 2023 ``Gravitational Wave Detection using Pulsar Timing Arrays". IC, LG and GT acknowledge financial support from the ``Action Th\'ematique de Cosmologie et Galaxies'' (ATCG), ``Action Th\'ematique Gravitation R\'ef\'erences Astronomie M\'etrologie'' (ATGRAM) and ``Action Th\'ematique Ph\'enomènes Extr\^emes et Multi-messagers'' (ATPEM) of CNRS/INSU, France.

Part of this work is based on observations
with the 100-m telescope of the Max-Planck-Institut für Radioastronomie (MPIfR) at Effelsberg in Germany. Pulsar research at the Jodrell Bank Centre for Astrophysics and the observations using the Lovell Telescope are supported by
a Consolidated Grant (ST/T000414/1) from the UK’s Science and Technology Facilities Council (STFC).
The Nan{\c c}ay radio Observatory is operated by the Paris Observatory, associated with the French Centre National de la Recherche Scientifique (CNRS), and partially supported by the Region Centre in France. The Westerbork Synthesis Radio Telescope is operated by the Netherlands Institute for Radio Astronomy (ASTRON) with support from the Netherlands Foundation for Scientific Research (NWO). The Sardinia Radio Telescope (SRT) is funded by the Department of University
and Research (MIUR), the Italian Space Agency (ASI), and the Autonomous Region of Sardinia (RAS) and is operated as a National Facility by the National Institute for Astrophysics (INAF). 

\appendix

\section{Analysis with \texttt{BAYESEPHEM} model}\label{sec:bayes}

We present results obtained using a more general model of SSE. We employ \texttt{BAYESEPHEM} \cite{NANOGrav:2020tig} to check for possible SSE systematics in the analysis.

This analysis represents a more conservative approach that can help mitigate the dipolar correlated signal induced by SSE systematics by a set of free parameters. However, \texttt{BAYESEPHEM} is known to partially absorb power from the GWB, reducing the evidence for the HD correlation in the EPTA data from 60 to 17 in terms of the Bayes factor~\cite{EPTA:2023fyk}. 

\begin{figure*}
    \centering
    \includegraphics[width=0.99\columnwidth]{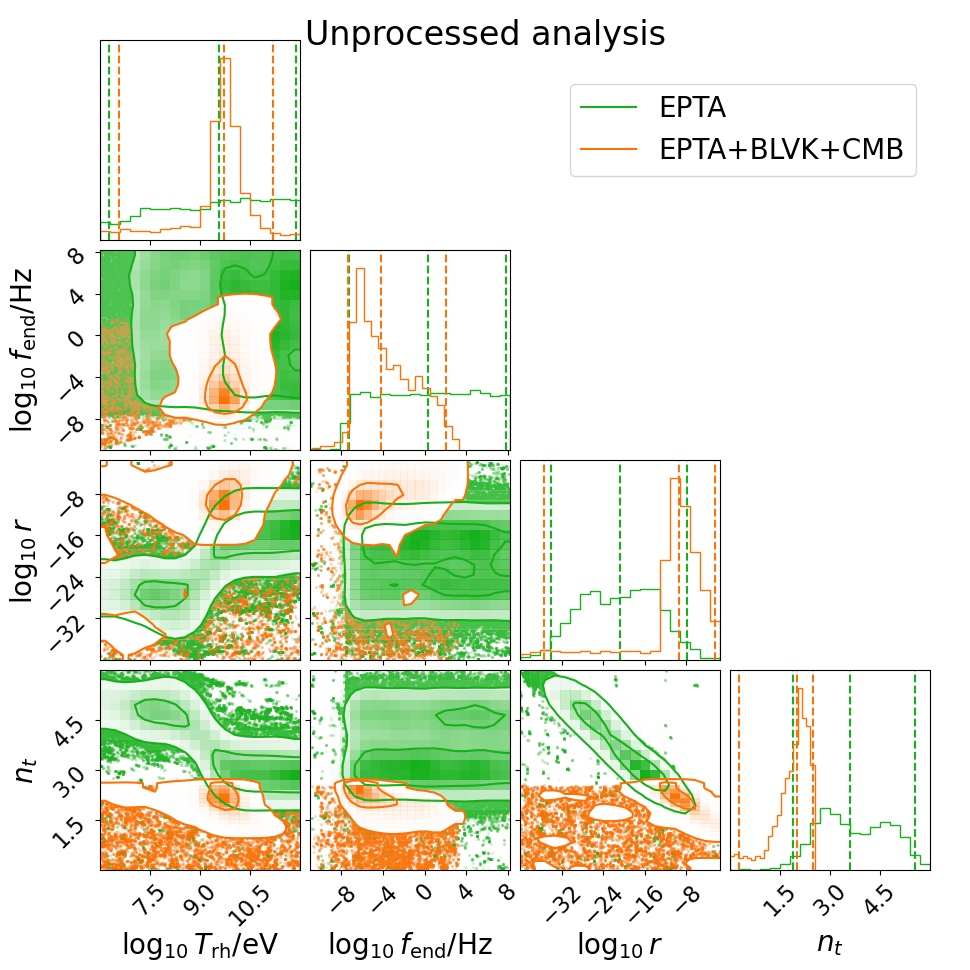}
    \includegraphics[width=0.99\columnwidth]{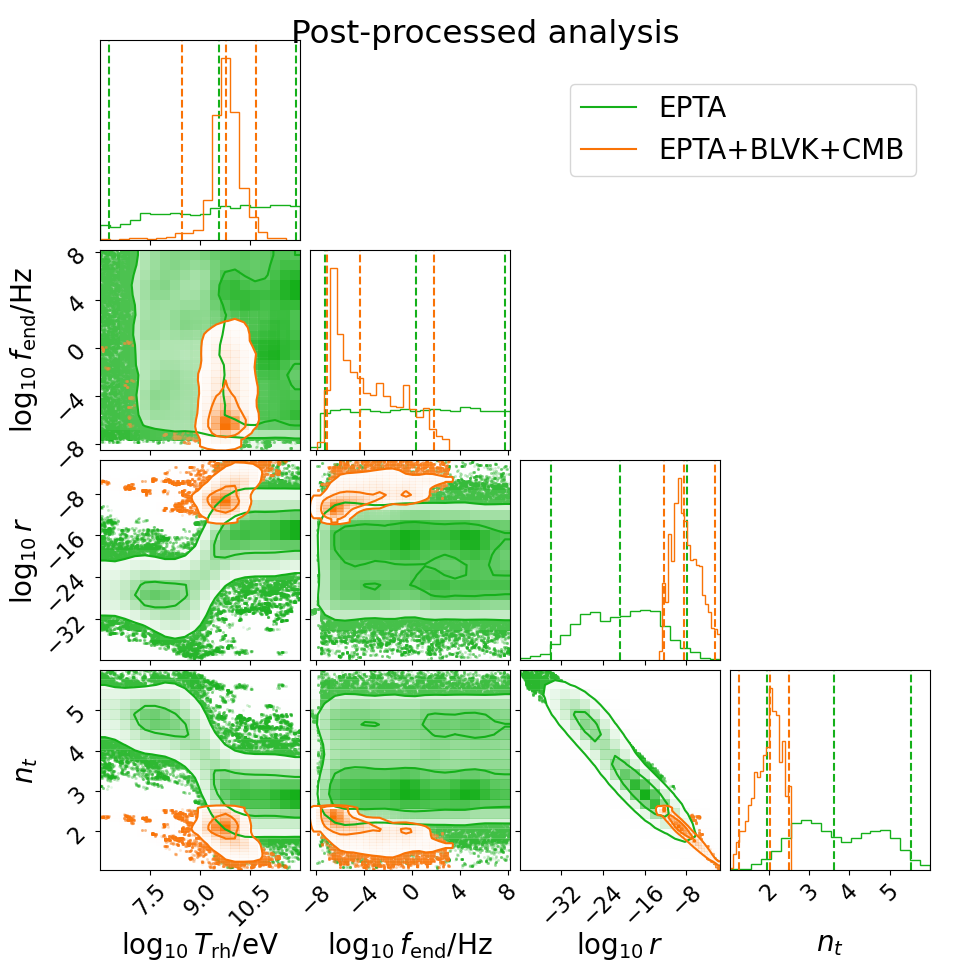}
    \caption{Posterior distributions for the $\Trh$, $\fend$, $r$ and $n_t$ parameters obtained from the EPTA (green) and EPTA+BLVK+CMB (orange) datasets. The smooth contours display the $1\sigma$ and $2\sigma$ credible regions of the posterior distributions. 
    {\it Left panel:} Parameter constraints obtained from the unprocessed chains. {\it Right panel:} Posterior distributions derived from the post-processing method. The SNR cut efficiently removes randomly distributed samples corresponding to a non-detection of CURN (see main text for details).}
    \label{fig:cosmoB}
\end{figure*}

The absorption of power from the GWB manifests as noise in Fig. \ref{fig:cosmoB}, which can be seen as randomly scattered points on the left 2-D corner plots. These broaden the posterior distributions and obscure the interpretation of the results.

\begin{figure*}[!t]
    \includegraphics[width=0.99\columnwidth]{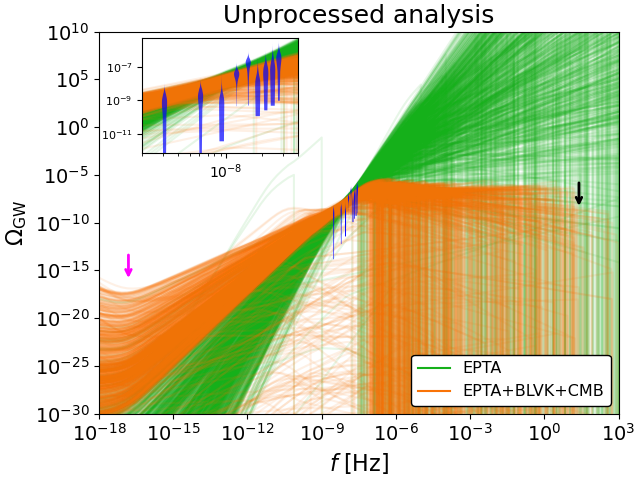}
    \includegraphics[width=0.99\columnwidth]{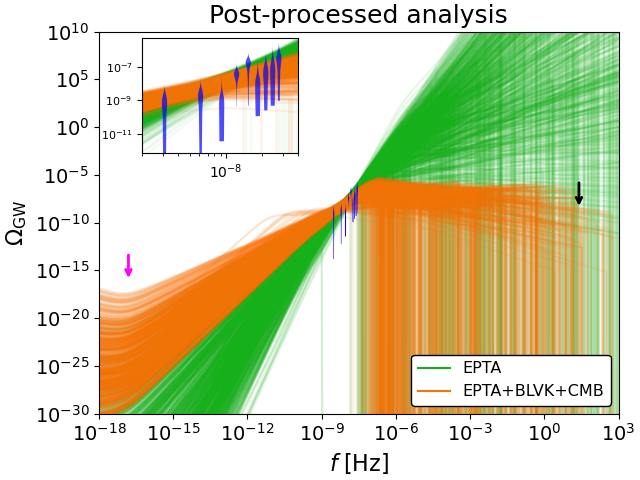}
    \caption{Theoretical predictions for $\Omega_{\rm GW}(f)$ randomly selected from the unprocessed ({\it left panel}) and post-processed ({\it right panel}) chains in the EPTA (green) and EPTA+BLVK+CMB (orange) data analyses. The shown models corresponds to a 1\% subsample of the respective chains. The blue violin plots represent the EPTA measurements obtained with the free spectrum method by sampling nine discrete EPTA frequencies~\cite{EPTA:2023fyk}. The downward black arrow indicates the $95\%$ upper limit on the GWB amplitude reported by the LIGO–Virgo–KAGRA (LVK) collaboration~\cite{LIGOScientific:2025bgj}, while the downward magenta arrow represents the CMB bound on $r$~\cite{Planck:2018jri}. In the unprocessed analysis, the curves that do not pass through the EPTA data points represent the noise. It can hence be seen that the post-processing method filters out these unwanted curves.}
    \label{fig:OmGWBB}
\end{figure*}

\begin{table*}[!t]
    \centering
    \begin{tabular}{lcccc}
    \hline\hline
    Data/Model 
    & $\log_{10}r$
    & $n_t$
    & $\enspace \log_{10}(\Trh/\eV)\enspace$
    & $\enspace \log_{10}(\fend/\Hz)\enspace$
    \\\hline
    {\bf Unprocessed}
    & 
    & 
    & 
    & 
    \\
    EPTA
    & $\enspace -20.69^{+12.90}_{-13.41}\enspace$
    & $\enspace 3.61^{+1.93}_{-1.73}\enspace$
    & $\enspace$Unconstrained$\enspace$
    & $\enspace > -7.26 \enspace$
    \\
    EPTA+BLVK
    &  $-15.53^{+11.92}_{-20.03}$
    &  $2.69^{+2.43}_{-2.12}$
    &  $ 9.44^{+1.43}_{-3.19}$
    &  $ -6.72^{+13.19}_{-0.88}$
    \\
    EPTA+BLVK+CMB
    & $-9.40^{+6.89}_{-26.08}$
    & $2.00^{+0.50}_{-1.73}$
    & $9.72^{+1.49}_{-3.13}$
    & $-4.18^{+6.25}_{-3.13}$
    \\\hline
    {\bf Post-processed}
    & 
    & 
    & 
    & 
    \\
    EPTA
    & $-20.65^{+12.82}_{-13.28}$
    & $3.62^{+1.92}_{-1.68}$
    & Unconstrained
    & $>-7.23$
    \\
    EPTA+BLVK
    &  $-12.80^{+9.59}_{-15.80}$
    &  $2.58^{+2.28}_{-1.24}$
    &  $9.62^{+0.95}_{-3.07}$
    &  $-6.67^{+12.45}_{-0.87}$
    \\
    EPTA+BLVK+CMB
    & $-8.43^{+6.03}_{-3.85}$
    & $2.02^{+0.47}_{-0.77}$
    & $9.78^{+0.90}_{-1.33}$
    & $-4.26^{+6.10}_{-2.75}$
    \\
    \hline
    \end{tabular}
    \caption{
    Mean and 95\% confidence intervals for the cosmological parameters in the EPTA, EPTA+BLVK and EPTA+BLVK+CMB analyses, derived from the unprocessed and post-processed chains. Excluding samples associated with a non-detection of CURN (post-processed) significantly reduce the error-bars in the EPTA+BLVK and EPTA+BLVK+CMB analysis, while it does not significantly affect the parameter constraints in the EPTA-only analysis. Two-dimensional posteriors are shown in Fig.~\ref{fig:cosmoB}.}
    \label{tab:param2}
\end{table*}

{\bf Post-processed analysis.} 
 To quantify the impact of these points on the cosmological constraints, we also report results from post-processed chains in which such samples are removed. In particular, we impose a lower bound on the signal-to-noise ratio (SNR) for the CURN detection, defined as 
\begin{equation}\label{eq:SNR}
    \frac{\mathcal{L}_{\rm full}(\theta_{\rm c},\,\theta_{{\rm PTA}})}{\mathcal{L}_{\rm no-CURN}(\theta_{{\rm PTA}})} \geq 10
\end{equation}
Here, $\mathcal{L}_{\rm full}(\theta_{\rm c},\theta_{{\rm PTA}})$ is the \texttt{ENTERPRISE} likelihood for the full model including a CURN component, and $\mathcal{L}_{\rm no-CURN}(\theta_{\rm PTA})$ is the likelihood of a model without the CURN component. The parameters $\theta_{\rm c}$ and $\theta_{{\rm PTA}}$ correspond to the cosmological parameters and the PTA parameters describing individual pulsars, respectively. Note that $\mathcal{L}_{\rm no-CURN}(\theta_{{\rm PTA}})$ is independent of the cosmological parameters, since the common red-noise amplitude is set to zero in this case. 

After applying the threshold ${\rm SNR}>10$, the sizes of the MCMC samples in the EPTA and EPTA+BLVK+CMB analyses are reduced by $1\%$ and $19\%$, respectively. Using a weaker criterion, ${\rm SNR}>1.01$, leads to the same fraction of discarded MCMC samples and does not affect our results. 

This indicates that after applying the SNR cut, the sampling noise in the EPTA+BLVK+CMB analysis is strongly suppressed, indicating that the randomly distributed points are indeed associated with scenarios in which the CURN signal is not detected. 
Applying the SNR cut consequently reduces the error-bars by $20\%-30\%$ and moderately shifts the posterior means when compared to the unprocessed EPTA+BLVK+CMB analysis, allowing for more informative results. It can also be understood that applying the SNR cut-off minimally affects the results for the unconstrained EPTA analysis. 
After applying this procedure, a consequence of the signal absorption induced by \texttt{BAYESEPHEM} can also be observed. This consequence is that the parameters associated with the tail belonging to the posterior of $\Trh$ at very low reheating temperatures in the EPTA+BLVK+CMB dataset is not distinguishable from the noise. 
Aside from the effect on the tail of $\Trh$, the results for EPTA+BLVK+CMB inferred from the post-processed chain using \texttt{BAYESEPHEM} are significant to the results obtained with {DE440}.
\bibliography{short.bib}

@article{McLaughlin:2013ira,
    author = "McLaughlin, Maura A.",
    title = "{The North American Nanohertz Observatory for Gravitational Waves}",
    eprint = "1310.0758",
    archivePrefix = "arXiv",
    primaryClass = "astro-ph.IM",
    doi = "10.1088/0264-9381/30/22/224008",
    journal = "Class. Quant. Grav.",
    volume = "30",
    pages = "224008",
    year = "2013"
}

@article{NANOGrav:2020bcs,
    author = "Arzoumanian, Zaven and others",
    collaboration = "NANOGrav",
    title = "{The NANOGrav 12.5 yr Data Set: Search for an Isotropic Stochastic Gravitational-wave Background}",
    eprint = "2009.04496",
    archivePrefix = "arXiv",
    primaryClass = "astro-ph.HE",
    doi = "10.3847/2041-8213/abd401",
    journal = "Astrophys. J. Lett.",
    volume = "905",
    number = "2",
    pages = "L34",
    year = "2020"
}

@article{ellis_jellis18ptmcmcsampler_2017_2,
	title = {jellis18/{PTMCMCSampler}: {Official} {Release}},
	shorttitle = {jellis18/{PTMCMCSampler}},
	url = {https://zenodo.org/records/1037579},
	doi = {10.5281/zenodo.1037579},
	urldate = {2026-01-22},
	publisher = {Zenodo},
	author = {Ellis, Justin and Haasteren, Rutger van},
	month = oct,
	year = {2017},
}

@article{NANOGrav:2023gor,
    author = "Agazie, Gabriella and others",
    collaboration = "NANOGrav",
    title = "{The NANOGrav 15 yr Data Set: Evidence for a Gravitational-wave Background}",
    eprint = "2306.16213",
    archivePrefix = "arXiv",
    primaryClass = "astro-ph.HE",
    doi = "10.3847/2041-8213/acdac6",
    journal = "Astrophys. J. Lett.",
    volume = "951",
    number = "1",
    pages = "L8",
    year = "2023"
}

@article{NANOGrav:2023hvm,
    author = "Afzal, Adeela and others",
    collaboration = "NANOGrav",
    title = "{The NANOGrav 15 yr Data Set: Search for Signals from New Physics}",
    eprint = "2306.16219",
    archivePrefix = "arXiv",
    primaryClass = "astro-ph.HE",
    reportNumber = "FERMILAB-PUB-23-589-T",
    doi = "10.3847/2041-8213/acdc91",
    journal = "Astrophys. J. Lett.",
    volume = "951",
    number = "1",
    pages = "L11",
    year = "2023",
    note = "[Erratum: Astrophys.J.Lett. 971, L27 (2024), Erratum: Astrophys.J. 971, L27 (2024)]"
}

@article{EPTA:2016ndq,
    author = "Desvignes, G. and others",
    collaboration = "EPTA",
    title = "{High-precision timing of 42 millisecond pulsars with the European Pulsar Timing Array}",
    eprint = "1602.08511",
    archivePrefix = "arXiv",
    primaryClass = "astro-ph.HE",
    doi = "10.1093/mnras/stw483",
    journal = "Mon. Not. Roy. Astron. Soc.",
    volume = "458",
    number = "3",
    pages = "3341--3380",
    year = "2016"
}

@Article{harris2020array,
 title         = {Array programming with {NumPy}},
 author        = {Charles R. Harris and K. Jarrod Millman and St{\'{e}}fan J.
                 van der Walt and Ralf Gommers and Pauli Virtanen and David
                 Cournapeau and Eric Wieser and Julian Taylor and Sebastian
                 Berg and Nathaniel J. Smith and Robert Kern and Matti Picus
                 and Stephan Hoyer and Marten H. van Kerkwijk and Matthew
                 Brett and Allan Haldane and Jaime Fern{\'{a}}ndez del
                 R{\'{i}}o and Mark Wiebe and Pearu Peterson and Pierre
                 G{\'{e}}rard-Marchant and Kevin Sheppard and Tyler Reddy and
                 Warren Weckesser and Hameer Abbasi and Christoph Gohlke and
                 Travis E. Oliphant},
 year          = {2020},
 month         = sep,
 journal       = {Nature},
 volume        = {585},
 number        = {7825},
 pages         = {357--362},
 doi           = {10.1038/s41586-020-2649-2},
 publisher     = {Springer Science and Business Media {LLC}},
 url           = {https://doi.org/10.1038/s41586-020-2649-2}
}

@article{EPTA:2021crs,
    author = "Chen, S. and others",
    collaboration = "EPTA",
    title = "{Common-red-signal analysis with 24-yr high-precision timing of the European Pulsar Timing Array: inferences in the stochastic gravitational-wave background search}",
    eprint = "2110.13184",
    archivePrefix = "arXiv",
    primaryClass = "astro-ph.HE",
    doi = "10.1093/mnras/stab2833",
    journal = "Mon. Not. Roy. Astron. Soc.",
    volume = "508",
    number = "4",
    pages = "4970--4993",
    year = "2021"
}

@article{EPTA:2023akd,
    author = "Antoniadis, J. and others",
    collaboration = "EPTA, InPTA",
    title = "{The second data release from the European Pulsar Timing Array - II. Customised pulsar noise models for spatially correlated gravitational waves}",
    eprint = "2306.16225",
    archivePrefix = "arXiv",
    primaryClass = "astro-ph.HE",
    doi = "10.1051/0004-6361/202346842",
    journal = "Astron. Astrophys.",
    volume = "678",
    pages = "A49",
    year = "2023"
}

@article{EPTA:2023fyk,
    author = "Antoniadis, J. and others",
    collaboration = "EPTA, InPTA:",
    title = "{The second data release from the European Pulsar Timing Array - III. Search for gravitational wave signals}",
    eprint = "2306.16214",
    archivePrefix = "arXiv",
    primaryClass = "astro-ph.HE",
    doi = "10.1051/0004-6361/202346844",
    journal = "Astron. Astrophys.",
    volume = "678",
    pages = "A50",
    year = "2023"
}

@article{EPTA:2023xxk,
    author = "Antoniadis, J. and others",
    collaboration = "EPTA, InPTA",
    title = "{The second data release from the European Pulsar Timing Array - IV. Implications for massive black holes, dark matter, and the early Universe}",
    eprint = "2306.16227",
    archivePrefix = "arXiv",
    primaryClass = "astro-ph.CO",
    doi = "10.1051/0004-6361/202347433",
    journal = "Astron. Astrophys.",
    volume = "685",
    pages = "A94",
    year = "2024"
}

@article{Antoniadis:2022pcn,
    author = "Antoniadis, J. and others",
    title = "{The International Pulsar Timing Array second data release: Search for an isotropic gravitational wave background}",
    eprint = "2201.03980",
    archivePrefix = "arXiv",
    primaryClass = "astro-ph.HE",
    doi = "10.1093/mnras/stab3418",
    journal = "Mon. Not. Roy. Astron. Soc.",
    volume = "510",
    number = "4",
    pages = "4873--4887",
    year = "2022"
}

@article{Caprini:2018mtu,
    author = "Caprini, Chiara and Figueroa, Daniel G.",
    title = "{Cosmological Backgrounds of Gravitational Waves}",
    eprint = "1801.04268",
    archivePrefix = "arXiv",
    primaryClass = "astro-ph.CO",
    doi = "10.1088/1361-6382/aac608",
    journal = "Class. Quant. Grav.",
    volume = "35",
    number = "16",
    pages = "163001",
    year = "2018"
}

@article{Caprini:2019egz,
    author = "Caprini, Chiara and others",
    title = "{Detecting gravitational waves from cosmological phase transitions with LISA: an update}",
    eprint = "1910.13125",
    archivePrefix = "arXiv",
    primaryClass = "astro-ph.CO",
    reportNumber = "DESY-19-159, IPPP/19/27, HIP-2019-14/TH, MITP/19-066, IFT-UAM/CSIC-19-139",
    doi = "10.1088/1475-7516/2020/03/024",
    journal = "JCAP",
    volume = "03",
    pages = "024",
    year = "2020"
}

@article{Allahverdi:2010xz,
    author = "Allahverdi, Rouzbeh and Brandenberger, Robert and Cyr-Racine, Francis-Yan and Mazumdar, Anupam",
    title = "{Reheating in Inflationary Cosmology: Theory and Applications}",
    eprint = "1001.2600",
    archivePrefix = "arXiv",
    primaryClass = "hep-th",
    doi = "10.1146/annurev.nucl.012809.104511",
    journal = "Ann. Rev. Nucl. Part. Sci.",
    volume = "60",
    pages = "27--51",
    year = "2010"
}

@article{Kuroyanagi:2008ye,
    author = "Kuroyanagi, Sachiko and Chiba, Takeshi and Sugiyama, Naoshi",
    title = "{Precision calculations of the gravitational wave background spectrum from inflation}",
    eprint = "0804.3249",
    archivePrefix = "arXiv",
    primaryClass = "astro-ph",
    doi = "10.1103/PhysRevD.79.103501",
    journal = "Phys. Rev. D",
    volume = "79",
    pages = "103501",
    year = "2009"
}

@article{Kuroyanagi:2014nba,
    author = "Kuroyanagi, Sachiko and Takahashi, Tomo and Yokoyama, Shuichiro",
    title = "{Blue-tilted Tensor Spectrum and Thermal History of the Universe}",
    eprint = "1407.4785",
    archivePrefix = "arXiv",
    primaryClass = "astro-ph.CO",
    reportNumber = "ICRR-REPORT-686-2014-12",
    doi = "10.1088/1475-7516/2015/02/003",
    journal = "JCAP",
    volume = "02",
    pages = "003",
    year = "2015"
}

@article{Kuroyanagi:2020sfw,
    author = "Kuroyanagi, Sachiko and Takahashi, Tomo and Yokoyama, Shuichiro",
    title = "{Blue-tilted inflationary tensor spectrum and reheating in the light of NANOGrav results}",
    eprint = "2011.03323",
    archivePrefix = "arXiv",
    primaryClass = "astro-ph.CO",
    doi = "10.1088/1475-7516/2021/01/071",
    journal = "JCAP",
    volume = "01",
    pages = "071",
    year = "2021"
}

@article{Vilenkin:1984ib,
    author = "Vilenkin, Alexander",
    title = "{Cosmic Strings and Domain Walls}",
    reportNumber = "PRINT-84-0840 (TUFTS)",
    doi = "10.1016/0370-1573(85)90033-X",
    journal = "Phys. Rept.",
    volume = "121",
    pages = "263--315",
    year = "1985"
}

@article{Hindmarsh:1994re,
    author = "Hindmarsh, M. B. and Kibble, T. W. B.",
    title = "{Cosmic strings}",
    eprint = "hep-ph/9411342",
    archivePrefix = "arXiv",
    reportNumber = "SUSX-TP-94-74, IMPERIAL-TP-94-95-5, NI-94025",
    doi = "10.1088/0034-4885/58/5/001",
    journal = "Rept. Prog. Phys.",
    volume = "58",
    pages = "477--562",
    year = "1995"
}

@article{Hindmarsh:2020hop,
    author = {Hindmarsh, Mark B. and L{\"u}ben, Marvin and Lumma, Johannes and Pauly, Martin},
    title = "{Phase transitions in the early universe}",
    eprint = "2008.09136",
    archivePrefix = "arXiv",
    primaryClass = "astro-ph.CO",
    reportNumber = "MPP-2020-163, HIP-2020-27/TH",
    doi = "10.21468/SciPostPhysLectNotes.24",
    journal = "SciPost Phys. Lect. Notes",
    volume = "24",
    pages = "1",
    year = "2021"
}

@article{Saikawa:2017hiv,
    author = "Saikawa, Ken'ichi",
    title = "{A review of gravitational waves from cosmic domain walls}",
    eprint = "1703.02576",
    archivePrefix = "arXiv",
    primaryClass = "hep-ph",
    reportNumber = "DESY-17-036",
    doi = "10.3390/universe3020040",
    journal = "Universe",
    volume = "3",
    number = "2",
    pages = "40",
    year = "2017"
}

@article{LIGOScientific:2025bgj,
    author = "Abac, A. G. and others",
    collaboration = "LIGO Scientific, VIRGO, KAGRA",
    title = "{Upper Limits on the Isotropic Gravitational-Wave Background from the first part of LIGO, Virgo, and KAGRA's fourth Observing Run}",
    eprint = "2508.20721",
    archivePrefix = "arXiv",
    primaryClass = "gr-qc",
    reportNumber = "LIGO-P2500349",
    month = "8",
    year = "2025"
}

@article{EuropeanPulsarTimingArray:2023lqe,
    author = "Quelquejay Leclere, Hippolyte and others",
    collaboration = "European Pulsar Timing Array, EPTA",
    title = "{Practical approaches to analyzing PTA data: Cosmic strings with six pulsars}",
    eprint = "2306.12234",
    archivePrefix = "arXiv",
    primaryClass = "gr-qc",
    doi = "10.1103/PhysRevD.108.123527",
    journal = "Phys. Rev. D",
    volume = "108",
    number = "12",
    pages = "123527",
    year = "2023"
}

@article{detweiler_pulsar_1979,
	title = {Pulsar timing measurements and the search for gravitational waves},
	volume = {234},
	issn = {0004-637X},
	url = {https://ui.adsabs.harvard.edu/abs/1979ApJ...234.1100D},
	doi = {10.1086/157593},
	urldate = {2026-02-15},
	journal = {The Astrophysical Journal},
	publisher = {IOP},
	author = {Detweiler, S.},
	month = dec,
	year = {1979},
	note = {ADS Bibcode: 1979ApJ...234.1100D},
	pages = {1100--1104},
}

@article{sazhin_opportunities_1978,
	title = {Opportunities for detecting ultralong gravitational waves},
	volume = {22},
	issn = {0038-5301},
	url = {https://ui.adsabs.harvard.edu/abs/1978SvA....22...36S},
	urldate = {2026-02-16},
	journal = {Soviet Astronomy},
	author = {Sazhin, M. V.},
	month = feb,
	year = {1978},
	note = {ADS Bibcode: 1978SvA....22...36S},
	pages = {36--38},
}

@article{Hellings:1983fr,
    author = "Hellings, R. w. and Downs, G. s.",
    title = "{UPPER LIMITS ON THE ISOTROPIC GRAVITATIONAL RADIATION BACKGROUND FROM PULSAR TIMING ANALYSIS}",
    doi = "10.1086/183954",
    journal = "Astrophys. J. Lett.",
    volume = "265",
    pages = "L39--L42",
    year = "1983"
}

@ARTICLE{1990ApJ...361..300F,
  author = {{Foster}, R.~S. and {Backer}, D.~C.},
  title = "{Constructing a Pulsar Timing Array}",
  journal = {\apj},
  year = {1990},
  volume = {361},
  pages = {300--308},
  doi = {10.1086/169195},
}

@article{Anber:2012du,
    author = "Anber, Mohamed M. and Sorbo, Lorenzo",
    title = "{Non-Gaussianities and chiral gravitational waves in natural steep inflation}",
    eprint = "1203.5849",
    archivePrefix = "arXiv",
    primaryClass = "astro-ph.CO",
    doi = "10.1103/PhysRevD.85.123537",
    journal = "Phys. Rev. D",
    volume = "85",
    pages = "123537",
    year = "2012"
}

@article{Cook:2011hg,
    author = "Cook, Jessica L. and Sorbo, Lorenzo",
    title = "{Particle production during inflation and gravitational waves detectable by ground-based interferometers}",
    eprint = "1109.0022",
    archivePrefix = "arXiv",
    primaryClass = "astro-ph.CO",
    doi = "10.1103/PhysRevD.85.023534",
    journal = "Phys. Rev. D",
    volume = "85",
    pages = "023534",
    year = "2012",
    note = "[Erratum: Phys.Rev.D 86, 069901 (2012)]"
}

@article{Namba:2015gja,
    author = "Namba, Ryo and Peloso, Marco and Shiraishi, Maresuke and Sorbo, Lorenzo and Unal, Caner",
    title = "{Scale-dependent gravitational waves from a rolling axion}",
    eprint = "1509.07521",
    archivePrefix = "arXiv",
    primaryClass = "astro-ph.CO",
    doi = "10.1088/1475-7516/2016/01/041",
    journal = "JCAP",
    volume = "01",
    pages = "041",
    year = "2016"
}

@article{Dimastrogiovanni:2016fuu,
    author = "Dimastrogiovanni, Emanuela and Fasiello, Matteo and Fujita, Tomohiro",
    title = "{Primordial Gravitational Waves from Axion-Gauge Fields Dynamics}",
    eprint = "1608.04216",
    archivePrefix = "arXiv",
    primaryClass = "astro-ph.CO",
    doi = "10.1088/1475-7516/2017/01/019",
    journal = "JCAP",
    volume = "01",
    pages = "019",
    year = "2017"
}

@article{Guzzetti:2016mkm,
    author = "Guzzetti, M. C. and Bartolo, N. and Liguori, M. and Matarrese, S.",
    title = "{Gravitational waves from inflation}",
    eprint = "1605.01615",
    archivePrefix = "arXiv",
    primaryClass = "astro-ph.CO",
    doi = "10.1393/ncr/i2016-10127-1",
    journal = "Riv. Nuovo Cim.",
    volume = "39",
    number = "9",
    pages = "399--495",
    year = "2016"
}

@article{Domenech:2021ztg,
    author = "Dom{\`e}nech, Guillem",
    title = "{Scalar Induced Gravitational Waves Review}",
    eprint = "2109.01398",
    archivePrefix = "arXiv",
    primaryClass = "gr-qc",
    doi = "10.3390/universe7110398",
    journal = "Universe",
    volume = "7",
    number = "11",
    pages = "398",
    year = "2021"
}

@article{Yuan:2021qgz,
    author = "Yuan, Chen and Huang, Qing-Guo",
    title = "{A topic review on probing primordial black hole dark matter with scalar induced gravitational waves}",
    eprint = "2103.04739",
    archivePrefix = "arXiv",
    primaryClass = "astro-ph.GA",
    doi = "10.1016/j.isci.2021.102860",
    journal = "iScience",
    volume = "24",
    pages = "102860",
    year = "2021"
}

@article{Caldwell:2017chz,
    author = "Caldwell, R. R. and Devulder, C.",
    title = "{Axion Gauge Field Inflation and Gravitational Leptogenesis: A Lower Bound on B Modes from the Matter-Antimatter Asymmetry of the Universe}",
    eprint = "1706.03765",
    archivePrefix = "arXiv",
    primaryClass = "astro-ph.CO",
    doi = "10.1103/PhysRevD.97.023532",
    journal = "Phys. Rev. D",
    volume = "97",
    number = "2",
    pages = "023532",
    year = "2018"
}

@article{Piao:2004tq,
    author = "Piao, Yun-Song and Zhang, Yuan-Zhong",
    title = "{Phantom inflation and primordial perturbation spectrum}",
    eprint = "astro-ph/0401231",
    archivePrefix = "arXiv",
    doi = "10.1103/PhysRevD.70.063513",
    journal = "Phys. Rev. D",
    volume = "70",
    pages = "063513",
    year = "2004"
}

@article{Kobayashi:2010cm,
    author = "Kobayashi, Tsutomu and Yamaguchi, Masahide and Yokoyama, Jun'ichi",
    title = "{G-inflation: Inflation driven by the Galileon field}",
    eprint = "1008.0603",
    archivePrefix = "arXiv",
    primaryClass = "hep-th",
    reportNumber = "RESCEU-18-10",
    doi = "10.1103/PhysRevLett.105.231302",
    journal = "Phys. Rev. Lett.",
    volume = "105",
    pages = "231302",
    year = "2010"
}

@article{Endlich:2012pz,
    author = "Endlich, Solomon and Nicolis, Alberto and Wang, Junpu",
    title = "{Solid Inflation}",
    eprint = "1210.0569",
    archivePrefix = "arXiv",
    primaryClass = "hep-th",
    doi = "10.1088/1475-7516/2013/10/011",
    journal = "JCAP",
    volume = "10",
    pages = "011",
    year = "2013"
}

@article{Fujita:2018ehq,
    author = "Fujita, Tomohiro and Kuroyanagi, Sachiko and Mizuno, Shuntaro and Mukohyama, Shinji",
    title = "{Blue-tilted Primordial Gravitational Waves from Massive Gravity}",
    eprint = "1808.02381",
    archivePrefix = "arXiv",
    primaryClass = "gr-qc",
    reportNumber = "YITP-18-84, IPMU18-0129",
    doi = "10.1016/j.physletb.2018.12.025",
    journal = "Phys. Lett. B",
    volume = "789",
    pages = "215--219",
    year = "2019"
}

@article{Planck:2018vyg,
    author = "Aghanim, N. and others",
    collaboration = "Planck",
    title = "{Planck 2018 results. VI. Cosmological parameters}",
    eprint = "1807.06209",
    archivePrefix = "arXiv",
    primaryClass = "astro-ph.CO",
    doi = "10.1051/0004-6361/201833910",
    journal = "Astron. Astrophys.",
    volume = "641",
    pages = "A6",
    year = "2020",
    note = "[Erratum: Astron.Astrophys. 652, C4 (2021)]"
}

@article{Planck:2018jri,
    author = "Akrami, Y. and others",
    collaboration = "Planck",
    title = "{Planck 2018 results. X. Constraints on inflation}",
    eprint = "1807.06211",
    archivePrefix = "arXiv",
    primaryClass = "astro-ph.CO",
    doi = "10.1051/0004-6361/201833887",
    journal = "Astron. Astrophys.",
    volume = "641",
    pages = "A10",
    year = "2020"
}

@article{Ellis:2020enterprise,
  author       = {Ellis, Justin A. and Vallisneri, Michele and Taylor, Stephen R. and Baker, Paul T.},
  title        = {{ENTERPRISE}: Enhanced Numerical Toolbox Enabling a Robust Pulsar Inference Suite},
  journal      = {Zenodo},
  year         = {2020},
  month        = {9},
  doi          = {10.5281/zenodo.4059815}
}

@article{Park_2021,
doi = {10.3847/1538-3881/abd414},
url = {https://doi.org/10.3847/1538-3881/abd414},
year = {2021},
month = {feb},
publisher = {The American Astronomical Society},
volume = {161},
number = {3},
pages = {105},
author = {Park, Ryan S. and Folkner, William M. and Williams, James G. and Boggs, Dale H.},
title = {The JPL Planetary and Lunar Ephemerides DE440 and DE441},
journal = {The Astronomical Journal},
}

@article{BICEP:2021xfz,
    author = "Ade, P. A. R. and others",
    collaboration = "BICEP, Keck",
    title = "{Improved Constraints on Primordial Gravitational Waves using Planck, WMAP, and BICEP/Keck Observations through the 2018 Observing Season}",
    eprint = "2110.00483",
    archivePrefix = "arXiv",
    primaryClass = "astro-ph.CO",
    doi = "10.1103/PhysRevLett.127.151301",
    journal = "Phys. Rev. Lett.",
    volume = "127",
    number = "15",
    pages = "151301",
    year = "2021"
}

@article{Balkenhol:2025wms,
    author = "Balkenhol, L. and others",
    title = "{Inflation at the End of 2025: Constraints on $r$ and $n_s$ Using the Latest CMB and BAO Data}",
    eprint = "2512.10613",
    archivePrefix = "arXiv",
    primaryClass = "astro-ph.CO",
    month = "12",
    year = "2025"
}

@article{NANOGrav:2020tig,
    author = "Vallisneri, M. and others",
    collaboration = "NANOGrav",
    title = "{Modeling the uncertainties of solar-system ephemerides for robust gravitational-wave searches with pulsar timing arrays}",
    eprint = "2001.00595",
    archivePrefix = "arXiv",
    primaryClass = "astro-ph.HE",
    doi = "10.3847/1538-4357/ab7b67",
    month = "1",
    year = "2020"
}

@article{Lewis:2019xzd,
    author = "Lewis, Antony",
    title = "{GetDist: a Python package for analysing Monte Carlo samples}",
    eprint = "1910.13970",
    archivePrefix = "arXiv",
    primaryClass = "astro-ph.IM",
    doi = "10.1088/1475-7516/2025/08/025",
    journal = "JCAP",
    volume = "08",
    pages = "025",
    year = "2025"
}

@article{Manchester:2012za,
    author = "Manchester, R. N. and others",
    title = "{The Parkes Pulsar Timing Array Project}",
    eprint = "1210.6130",
    archivePrefix = "arXiv",
    primaryClass = "astro-ph.IM",
    doi = "10.1017/pasa.2012.017",
    journal = "Publ. Astron. Soc. Austral.",
    volume = "30",
    pages = "17",
    year = "2013"
}

@article{Goncharov:2021oub,
    author = "Goncharov, Boris and others",
    title = "{On the Evidence for a Common-spectrum Process in the Search for the Nanohertz Gravitational-wave Background with the Parkes Pulsar Timing Array}",
    eprint = "2107.12112",
    archivePrefix = "arXiv",
    primaryClass = "astro-ph.HE",
    doi = "10.3847/2041-8213/ac17f4",
    journal = "Astrophys. J. Lett.",
    volume = "917",
    number = "2",
    pages = "L19",
    year = "2021"
}

@article{Turner:1993vb,
    author = "Turner, Michael S. and White, Martin J. and Lidsey, James E.",
    title = "{Tensor perturbations in inflationary models as a probe of cosmology}",
    eprint = "astro-ph/9306029",
    archivePrefix = "arXiv",
    reportNumber = "FERMILAB-PUB-93-069-A, CFPA-TH-93-19, CFPA-93-19",
    doi = "10.1103/PhysRevD.48.4613",
    journal = "Phys. Rev. D",
    volume = "48",
    pages = "4613--4622",
    year = "1993"
}

@article{Laine:2015kra,
    author = "Laine, M. and Meyer, M.",
    title = "{Standard Model thermodynamics across the electroweak crossover}",
    eprint = "1503.04935",
    archivePrefix = "arXiv",
    primaryClass = "hep-ph",
    doi = "10.1088/1475-7516/2015/07/035",
    journal = "JCAP",
    volume = "07",
    pages = "035",
    year = "2015"
}

@article{Lyth:1995ka,
    author = "Lyth, David H. and Stewart, Ewan D.",
    title = "{Thermal inflation and the moduli problem}",
    eprint = "hep-ph/9510204",
    archivePrefix = "arXiv",
    reportNumber = "LANCS-TH-9505, RESCEU-16-95, LANCASTER-TH-9505",
    doi = "10.1103/PhysRevD.53.1784",
    journal = "Phys. Rev. D",
    volume = "53",
    pages = "1784--1798",
    year = "1996"
}

@article{Brandenberger:2016vhg,
    author = "Brandenberger, Robert and Peter, Patrick",
    title = "{Bouncing Cosmologies: Progress and Problems}",
    eprint = "1603.05834",
    archivePrefix = "arXiv",
    primaryClass = "hep-th",
    doi = "10.1007/s10701-016-0057-0",
    journal = "Found. Phys.",
    volume = "47",
    number = "6",
    pages = "797--850",
    year = "2017"
}

@article{Lasky:2015lej,
    author = "Lasky, Paul D. and others",
    title = "{Gravitational-wave cosmology across 29 decades in frequency}",
    eprint = "1511.05994",
    archivePrefix = "arXiv",
    primaryClass = "astro-ph.CO",
    doi = "10.1103/PhysRevX.6.011035",
    journal = "Phys. Rev. X",
    volume = "6",
    number = "1",
    pages = "011035",
    year = "2016"
}

@article{ForemanMackey:2016corner,
  author         = "Foreman-Mackey, Daniel",
  title          = "{corner.py: Scatterplot matrices in Python}",
  journal        = "J. Open Source Softw.",
  volume         = "1",
  number         = "2",
  pages          = "24",
  year           = "2016",
  month          = "6",
  doi            = "10.21105/joss.00024",
  url            = "https://doi.org/10.21105/joss.00024"
}

@article{Cornish:2013aba,
    author = "Cornish, Neil J. and Sesana, A.",
    title = "{Pulsar Timing Array Analysis for Black Hole Backgrounds}",
    eprint = "1305.0326",
    archivePrefix = "arXiv",
    primaryClass = "gr-qc",
    doi = "10.1088/0264-9381/30/22/224005",
    journal = "Class. Quant. Grav.",
    volume = "30",
    pages = "224005",
    year = "2013"
}

@book{Maggiore:2007ulw,
    author = "Maggiore, Michele",
    title = "{Gravitational Waves. Vol. 1: Theory and Experiments}",
    doi = "10.1093/acprof:oso/9780198570745.001.0001",
    isbn = "978-0-19-171766-6, 978-0-19-852074-0",
    publisher = "Oxford University Press",
    year = "2007"
}

@article{Planck:2013jfk,
    author = "Ade, P. A. R. and others",
    collaboration = "Planck",
    title = "{Planck 2013 results. XXII. Constraints on inflation}",
    eprint = "1303.5082",
    archivePrefix = "arXiv",
    primaryClass = "astro-ph.CO",
    reportNumber = "CERN-PH-TH-2013-135",
    doi = "10.1051/0004-6361/201321569",
    journal = "Astron. Astrophys.",
    volume = "571",
    pages = "A22",
    year = "2014"
}

@article{Vagnozzi:2023lwo,
    author = "Vagnozzi, Sunny",
    title = "{Inflationary interpretation of the stochastic gravitational wave background signal detected by pulsar timing array experiments}",
    eprint = "2306.16912",
    archivePrefix = "arXiv",
    primaryClass = "astro-ph.CO",
    doi = "10.1016/j.jheap.2023.07.001",
    journal = "JHEAp",
    volume = "39",
    pages = "81--98",
    year = "2023"
}

@article{Figueroa:2023zhu,
    author = "Figueroa, Daniel G. and Pieroni, Mauro and Ricciardone, Angelo and Simakachorn, Peera",
    title = "{Cosmological Background Interpretation of Pulsar Timing Array Data}",
    eprint = "2307.02399",
    archivePrefix = "arXiv",
    primaryClass = "astro-ph.CO",
    reportNumber = "CERN-TH-2023-132",
    doi = "10.1103/PhysRevLett.132.171002",
    journal = "Phys. Rev. Lett.",
    volume = "132",
    number = "17",
    pages = "171002",
    year = "2024"
}

@article{Balaji:2023ehk,
    author = "Balaji, Shyam and Dom{\`e}nech, Guillem and Franciolini, Gabriele",
    title = "{Scalar-induced gravitational wave interpretation of PTA data: the role of scalar fluctuation propagation speed}",
    eprint = "2307.08552",
    archivePrefix = "arXiv",
    primaryClass = "gr-qc",
    doi = "10.1088/1475-7516/2023/10/041",
    journal = "JCAP",
    volume = "10",
    pages = "041",
    year = "2023"
}

@article{Franciolini:2023pbf,
    author = "Franciolini, Gabriele and Iovino, Junior., Antonio and Vaskonen, Ville and Veermae, Hardi",
    title = "{Recent Gravitational Wave Observation by Pulsar Timing Arrays and Primordial Black Holes: The Importance of Non-Gaussianities}",
    eprint = "2306.17149",
    archivePrefix = "arXiv",
    primaryClass = "astro-ph.CO",
    doi = "10.1103/PhysRevLett.131.201401",
    journal = "Phys. Rev. Lett.",
    volume = "131",
    number = "20",
    pages = "201401",
    year = "2023"
}

@article{RoperPol:2022iel,
    author = "Roper Pol, Alberto and Caprini, Chiara and Neronov, Andrii and Semikoz, Dmitri",
    title = "{Gravitational wave signal from primordial magnetic fields in the Pulsar Timing Array frequency band}",
    eprint = "2201.05630",
    archivePrefix = "arXiv",
    primaryClass = "astro-ph.CO",
    doi = "10.1103/PhysRevD.105.123502",
    journal = "Phys. Rev. D",
    volume = "105",
    number = "12",
    pages = "123502",
    year = "2022"
}

@article{Yin:2024ccm,
    author = "Yin, Lu",
    title = "{Does Gauss-Bonnet inflationary gravitational waves satisfy the pulsar timing arrays observations?}",
    eprint = "2410.07949",
    archivePrefix = "arXiv",
    primaryClass = "astro-ph.CO",
    doi = "10.1088/1475-7516/2025/05/047",
    journal = "JCAP",
    volume = "05",
    pages = "047",
    year = "2025"
}

@article{Bernardo:2025lie,
    author = "Bernardo, Reginald Christian and Koh, Seoktae and Tumurtushaa, Gansukh",
    title = "{Implications of Pulsar Timing Arrays for Gauss-Bonnet inflation}",
    eprint = "2505.10235",
    archivePrefix = "arXiv",
    primaryClass = "astro-ph.CO",
    doi = "10.1088/1475-7516/2025/10/013",
    journal = "JCAP",
    volume = "10",
    pages = "013",
    year = "2025"
}

@article{Cheung:2007st,
    author = "Cheung, Clifford and Creminelli, Paolo and Fitzpatrick, A. Liam and Kaplan, Jared and Senatore, Leonardo",
    title = "{The Effective Field Theory of Inflation}",
    eprint = "0709.0293",
    archivePrefix = "arXiv",
    primaryClass = "hep-th",
    reportNumber = "IC-2007-032",
    doi = "10.1088/1126-6708/2008/03/014",
    journal = "JHEP",
    volume = "03",
    pages = "014",
    year = "2008"
}

@article{Bianchi:2024qyp,
    author = "Bianchi, Eugenio and Gamonal, Mauricio",
    title = "{Primordial power spectrum at N3LO in effective theories of inflation}",
    eprint = "2405.03157",
    archivePrefix = "arXiv",
    primaryClass = "gr-qc",
    doi = "10.1103/PhysRevD.110.104032",
    journal = "Phys. Rev. D",
    volume = "110",
    number = "10",
    pages = "104032",
    year = "2024"
}

@article{Kuroyanagi:2014qza,
    author = "Kuroyanagi, Sachiko and Nakayama, Kazunori and Yokoyama, Jun'ichi",
    title = "{Prospects of determination of reheating temperature after inflation by DECIGO}",
    eprint = "1410.6618",
    archivePrefix = "arXiv",
    primaryClass = "astro-ph.CO",
    reportNumber = "RESCEU-44-14",
    doi = "10.1093/ptep/ptu176",
    journal = "PTEP",
    volume = "2015",
    number = "1",
    pages = "013E02",
    year = "2015"
}

@article{Kuroyanagi:2010mm,
    author = "Kuroyanagi, Sachiko and Chiba, Takeshi and Sugiyama, Naoshi",
    title = "{Prospects for Direct Detection of Inflationary Gravitational Waves by Next Generation Interferometric Detectors}",
    eprint = "1010.5246",
    archivePrefix = "arXiv",
    primaryClass = "astro-ph.CO",
    reportNumber = "ICRR-REPORT-575-2010-8",
    doi = "10.1103/PhysRevD.83.043514",
    journal = "Phys. Rev. D",
    volume = "83",
    pages = "043514",
    year = "2011"
}

@article{Elbers:2025vlz,
    author = "Elbers, W. and others",
    title = "{Constraints on neutrino physics from DESI DR2 BAO and DR1 full shape}",
    eprint = "2503.14744",
    archivePrefix = "arXiv",
    primaryClass = "astro-ph.CO",
    reportNumber = "FERMILAB-PUB-25-0168-PPD",
    doi = "10.1103/w9pk-xsk7",
    journal = "Phys. Rev. D",
    volume = "112",
    number = "8",
    pages = "083513",
    year = "2025"
}

@article{AtacamaCosmologyTelescope:2025nti,
    author = "Calabrese, Erminia and others",
    collaboration = "Atacama Cosmology Telescope",
    title = "{The Atacama Cosmology Telescope: DR6 constraints on extended cosmological models}",
    eprint = "2503.14454",
    archivePrefix = "arXiv",
    primaryClass = "astro-ph.CO",
    reportNumber = "FERMILAB-PUB-25-0157-PPD",
    doi = "10.1088/1475-7516/2025/11/063",
    journal = "JCAP",
    volume = "11",
    pages = "063",
    year = "2025"
}

@article{Cielo:2023bqp,
    author = "Cielo, Mattia and Escudero, Miguel and Mangano, Gianpiero and Pisanti, Ofelia",
    title = "{Neff in the Standard Model at NLO is 3.043}",
    eprint = "2306.05460",
    archivePrefix = "arXiv",
    primaryClass = "hep-ph",
    reportNumber = "CERN-TH-2023-103",
    doi = "10.1103/PhysRevD.108.L121301",
    journal = "Phys. Rev. D",
    volume = "108",
    number = "12",
    pages = "L121301",
    year = "2023"
}

@article{Schoneberg:2024ifp,
    author = {Sch{\"o}neberg, Nils},
    title = "{The 2024 BBN baryon abundance update}",
    eprint = "2401.15054",
    archivePrefix = "arXiv",
    primaryClass = "astro-ph.CO",
    doi = "10.1088/1475-7516/2024/06/006",
    journal = "JCAP",
    volume = "06",
    pages = "006",
    year = "2024"
}

@article{Garcia:2020wiy,
    author = "Garcia, Marcos A. G. and Kaneta, Kunio and Mambrini, Yann and Olive, Keith A.",
    title = "{Inflaton Oscillations and Post-Inflationary Reheating}",
    eprint = "2012.10756",
    archivePrefix = "arXiv",
    primaryClass = "hep-ph",
    reportNumber = "UMN-TH-4006/20, FTPI-MINN-20/37, IFT-UAM/CSIC-20-185, KIAS-P20071",
    doi = "10.1088/1475-7516/2021/04/012",
    journal = "JCAP",
    volume = "04",
    pages = "012",
    year = "2021"
}

@article{Ahmed:2021fvt,
    author = "Ahmed, Aqeel and Grzadkowski, Bohdan and Socha, Anna",
    title = "{Implications of time-dependent inflaton decay on reheating and dark matter production}",
    eprint = "2111.06065",
    archivePrefix = "arXiv",
    primaryClass = "hep-ph",
    doi = "10.1016/j.physletb.2022.137201",
    journal = "Phys. Lett. B",
    volume = "831",
    pages = "137201",
    year = "2022"
}

@article{Drewes:2014pfa,
    author = "Drewes, Marco",
    title = "{On finite density effects on cosmic reheating and moduli decay and implications for Dark Matter production}",
    eprint = "1406.6243",
    archivePrefix = "arXiv",
    primaryClass = "hep-ph",
    reportNumber = "TUM-HEP-949-14, NSF-KITP-14-059",
    doi = "10.1088/1475-7516/2014/11/020",
    journal = "JCAP",
    volume = "11",
    pages = "020",
    year = "2014"
}
\end{document}